\documentclass[letterpaper,11pt]{article}

\usepackage{arydshln}
\usepackage{amsmath}
\usepackage{amsthm}
\usepackage{authblk}
\usepackage{setspace}
\usepackage{url}
\usepackage[comma,authoryear]{natbib}
\usepackage{hyperref}
\usepackage{theoremref}
\usepackage[titletoc,title]{appendix}
\usepackage[table]{xcolor}
\usepackage{float}
\restylefloat{table}
\usepackage{amssymb}
\usepackage{caption}
\usepackage{subcaption}
\usepackage[shortlabels]{enumitem}
\setcitestyle{citesep={;}}

\definecolor{armygreen}{rgb}{0.29, 0.33, 0.13}
\definecolor{auburn}{rgb}{0.43, 0.21, 0.1}
\definecolor{burgundy}{rgb}{0.5, 0.0, 0.13}
\definecolor{medium red}{rgb}{.490,.298,.337}
\definecolor{dark red}{rgb}{.235,.141,.161}
\definecolor{dark green}{rgb}{0.0,0.5,0.0}

\hypersetup{
	colorlinks = true,
	linkcolor = {burgundy},
	urlcolor = {burgundy},
	citecolor = {burgundy}, 		
	linkbordercolor = {white},
}

\newtheorem{theorem}{Theorem}
\newtheorem{theorem*}{Theorem*}
\newtheorem{proposition}{Proposition}
\newtheorem{proposition*}{Proposition*}
\newtheorem{claim}{Claim}[section]
\newtheorem{lemma}{Lemma}[section]
\newtheorem{corollary}{Corollary}
\newtheorem{corollary*}{Corollary*}

\theoremstyle{definition}
\newtheorem{definition}{Definition}
\theoremstyle{definition}
\newtheorem{example}{Example}
\theoremstyle{definition}
\newtheorem{remark}{Remark}
\theoremstyle{definition}
\newtheorem{note}{Note}

\newcommand*{\claimproofname}{Proof of Claim}
\newenvironment{claimproof}[1][\claimproofname]{\begin{proof}[#1]}{\end{proof}}

\usepackage{verbatim}
\usepackage{tikz}
\tikzset{
	treenode/.style = {shape=rectangle, rounded corners,
		draw, align=center,
		top color=white, bottom color=blue!20},
	root/.style = {shape=rectangle, rounded corners,
		draw, align=center,
		top color=white, bottom color=blue!20},
	root1/.style = {shape=rectangle, rounded corners,
		draw, align=center,
		top color=white, bottom color=red!40},
	root2/.style = {shape=rectangle, rounded corners,
		draw, align=center,
		top color=white, bottom color=red!20},
	env/.style = {shape=rectangle, rounded corners,
		draw, align=center,
		top color=white, bottom color=blue!20},
}

\title{Compatibility between Stability and Strategy-Proofness: A Single-Peaked Preferences Investigation\thanks{An earlier version of this paper circulated under the title ``\textit{Compatibility between stability and strategy-proofness with single-peaked preferences on trees}''. I am grateful to Szilvia P\'{a}pai for her insightful feedback. I also thank Hector Chade, Andrew Mackenzie, and Vikram Manjunath for their helpful suggestions. I further appreciate the valuable comments of the editor, Nicholas Yannelis, the associate editor, and two anonymous referees.}}

\author{Pinaki Mandal\thanks{E-mail: \textit{pinaki.mandal@asu.edu}}}

\affil{Department of Economics, Arizona State University, Tempe, USA}

\date{May 8, 2025}

\begin{document} 	
	\maketitle
	
	\begin{abstract}
		In two-sided matching markets, ensuring both stability and strategy-proofness poses a significant challenge; it is impossible when agents' preferences are unrestricted. But what if agents' preferences have specific restricted structures? Such scenarios frequently arise in real-world applications. This study explores the possibility of achieving both stability and strategy-proofness by focusing on scenarios where agents' preferences follow a structured pattern called \textit{single-peakedness}.
		
		We focus on the simplest case -- the well-known \textit{marriage problem}, which is a one-to-one matching market. Despite its simplicity, this model is a useful starting point for exploration in many cases. Our main contribution is identifying all single-peaked subdomains on which stability and (weak/strong group) strategy-proofness are compatible, which we present through two key results. The first one characterizes all single-peaked subdomains with stable and (weakly group) strategy-proof matching rules, and identifies such a matching rule on these domains. The second one is an impossibility result that shows the incompatibility between stability and strong group strategy-proofness on single-peaked subdomains.
	\end{abstract}
	
	

	\newpage

	\section{Introduction}

	The theory of two-sided matching markets is highly relevant to the design of many real-world institutions, such as assigning graduates to residency programs, students to colleges, or workers to firms. Such markets consist of two groups of agents, with each agent on one side having a (strict) preference over the agents on the other side. A matching is selected based on the agents' preferences, where each agent is matched with a subset of agents on the other side in a way such that if an agent is in another agent's match, then the second agent must be in the first agent's match. Although not all two-sided matching markets can be described precisely in these terms, the model we analyze is simple yet meaningful enough to serve as a starting point for exploration in many cases.
	
	In such markets, \textit{stability} \citep{gale1962college} has been considered a desirable property to be satisfied by any matching.\footnote{In real-world applications, empirical studies have shown that stable mechanisms often succeed, while unstable ones fail. See \citet{roth2002economist} for a summary of this evidence.} A matching is stable if no pair of agents -- one on each side -- would rather be matched to each other than to one of their current matches. Apart from stability, it is also crucial to ensure that the incentives are aligned when participating agents are strategic. The standard notion of \textit{strategy-proofness} requires truth-telling as a dominant strategy for every agent. Unfortunately, achieving stability and strategy-proofness simultaneously is impossible when the agents' preferences are unrestricted (see, for example, \citet{roth1982economics}).
	
	What if agents' preferences have specific restricted structures? We often encounter such situations in real-world scenarios, and it is natural to wonder whether stability and strategy-proofness can be achieved simultaneously in such situations. This paper provides a comprehensive solution to this question when agents have \textit{single-peaked preferences} \citep{black1948rationale}. Single-peakedness is a natural restriction on the preferences, where agents on each side are ordered based on specific criteria (potentially different criteria for each side), and the preferences respect these orderings in the sense that as an agent moves away from their most preferred choice, their welfare declines.
	
	For our analysis, we focus on the simplest case -- the well-known \textit{marriage problem} \citep{gale1962college}, which is a one-to-one matching market. As is the convention, we refer to the two sides of this market as ``men'' and ``women''. We further assume that the market is balanced (i.e., an equal number of men and women), and every agent prefers any match to remain unmatched.

	\subsection{Overview of our results}

	Before delving into our results, we introduce two mild conditions that we impose on the domains. The first one is called \textit{richness}, which requires that for every man and every woman, the man has an admissible preference that ranks the woman first, and the woman has an admissible preference that ranks the man first. This condition reflects the idea that every agent should be allowed to express a strong preference for any potential match, ensuring flexibility in individual preferences. The second domain condition, \textit{anonymity}, is based on the idea of a common-domain approach. It requires every man to have the same set of admissible preferences, and so do the women. This symmetry ensures that agents within each group (men or women) are treated uniformly in terms of the options available to them. All of our results assume the domain satisfies these two conditions.
	
	As mentioned earlier, we explore the compatibility between stability and strategy-proofness when agents have single-peaked preferences. It is important to note that our focus is not limited to the \textit{maximal single-peaked domain}; instead, we are considering all rich and anonymous subdomains of the maximal single-peaked domain, including the domain itself.
	
	Our first result (Theorem \ref{theo sp-wgsp classical single peaked}) characterizes all single-peaked subdomains with stable and strategy-proof matching rules. Such domains must satisfy a restriction, called \textit{top dominance} \citep{alcalde1994top}, for at least one side of the market. Furthermore, on these domains, the \textit{deferred acceptance (DA) rule} \citep{gale1962college} -- a well-known stable matching rule -- is strategy-proof (the domain's structure determines the DA's proposing side).\footnote{The DA rule is also widely used in \textit{priority-based allocation problems}, where its incentive properties have been extensively analyzed. See, for instance, \citet{ergin2002efficient}, \citet{kojima2011robust}, \citet{afacan2012group}, and \citet{han2025characterizing}.} 
	Top dominance, which is the required domain restriction for our characterization result, roughly requires that for every triplet of outcomes for an agent, if there exists an admissible preference that favors the first outcome over the second and the second over the third, then there cannot be another admissible preference that favors the first outcome over the third and the third over the second.
	
	Since satisfying top dominance (for at least one side of the market) is both a necessary and sufficient condition for a single-peaked subdomain to have stable and strategy-proof matching rules (as demonstrated by Theorem \ref{theo sp-wgsp classical single peaked}), it is natural to ask what types of domain structures emerge within the single-peaked subdomains under top dominance. One example is when the sets of admissible preferences consist of \textit{left-biased single-peaked preferences} \citep{alva2017manipulation}. We provide a detailed discussion with additional examples of such domains in Appendix \ref{appendix domain structure}.
	
	As a corollary of Theorem \ref{theo sp-wgsp classical single peaked}, we establish that stability and strategy-proofness are incompatible on the maximal single-peaked domain (Corollary \ref{corollary impossibility on maximal single peaked}). To understand why this is the case, note that the maximal single-peaked domain does not satisfy top dominance for either side of the market, preventing it from meeting the necessary condition to admit a stable and strategy-proof matching rule.
	
	Since stability and strategy-proofness can be achieved simultaneously on certain single-peaked subdomains (as demonstrated by Theorem \ref{theo sp-wgsp tree single peaked}), it is natural to ask whether stability can coexist with stronger incentive properties on such domains. In this paper, we present two such notions -- \textit{weak group strategy-proofness} and \textit{strong group strategy-proofness}, and explore their compatibility with stability on single-peaked subdomains. 
	Weak group strategy-proofness ensures that no coalition of agents can be strictly better off by misreporting their preferences. On the other hand, strong group strategy-proofness ensures that no coalition of agents can be weakly better off by misreporting their preferences, with some agents in the coalition strictly better off. It is worth noting that a coalition may consist of both men and women.
	
	We demonstrate that the single-peaked subdomains on which stability and weak group strategy-proofness are compatible are the same ones on which stability and strategy-proofness are compatible (Theorem \ref{theo sp-wgsp classical single peaked}). In fact, on every such domain, the DA rule is weakly group strategy-proof (the domain's structure determines the DA's proposing side). This insight helps to explain why the DA rule is often favored in real-world applications. When stability and strategy-proofness can be achieved simultaneously on a single-peaked subdomain, the DA rule becomes the natural choice, offering not only these desirable properties but also resistance to group manipulation.
	
	However, unlike weak group strategy-proofness, strong group strategy-proofness cannot be achieved together with stability on single-peaked subdomains. Specifically, whenever either side of the market has at least five agents, no stable matching rule on a single-peaked subdomain is strongly group strategy-proof (Theorem \ref{theorem no stable and gsp}). In fact, we prove a stronger result (Proposition \ref{proposition no stable and nb rule for atleast 5}), which shows that whenever either side has at least five agents, there is no single-peaked subdomain with a stable and \textit{non-bossy} \citep{satterthwaite1981strategy} matching rule. Non-bossiness is a weaker notion than strong group strategy-proofness, which requires that agents can only change others' matches if they change their own.
	
	So far, we have focused on \textit{classical single-peaked preferences}, where single-peakedness is defined with respect to a \textit{linear} prior ordering over the alternatives. However, many real-world decision-making scenarios are more intricate: alternatives are often interconnected as nodes within a network, rather than arranged along a simple linear spectrum. In these cases, agents' preferences are naturally shaped by the structure of the underlying network.
	For example, imagine a scenario where all the agents are located on a ``road network'', and their welfare diminishes with increasing distance. In such cases, a man will prefer to be matched with a woman over another woman if the first woman is located on every possible road between the man and the other woman. This illustrates how the network influences preference orderings, requiring us to account for topological constraints in our setting.
	
	The discussion above emphasizes the importance of moving beyond classical single-peakedness to a network-based notion of single-peakedness. Specifically, we focus on networks with a \textit{tree} structure, which provides a natural extension of the linear model. Trees retain many desirable properties of linear orderings while simplifying the analysis of network-based preferences. We introduce the concept of \textit{single-peakedness on trees} \citep{demange1982single} as a generalization of classical single-peakedness, and discuss the possibility of extending our results to this framework. We demonstrate that all the results established for classical single-peaked subdomains also hold for \textit{tree-single-peaked subdomains}.
	
	To summarize, our main contribution in this paper is identifying all (tree-)single-peaked subdomains on which stability and (weak/strong group) strategy-proofness are compatible, which we present through two key results. The first one (Theorem$^*$ \ref{theo sp-wgsp tree single peaked}) characterizes all (tree-)single-peaked subdomains with stable and (weakly group) strategy-proof matching rules, and identifies such a matching rule on these domains. The second one (Theorem$^*$ \ref{theorem no stable and gsp tree}) is an impossibility result that shows the incompatibility between stability and strong group strategy-proofness on (tree-)single-peaked subdomains. 
	It is important to note that while our first result establishes that the (tree-)single-peaked subdomains on which stability and strategy-proofness are compatible are the same ones on which stability and weak group strategy-proofness are compatible, it does not imply that every stable and strategy-proof matching rule on such a domain is weakly group strategy-proof.\footnote{Strategy-proof rules often satisfy weak group strategy-proofness under suitable domain conditions; see \citet{le2009equivalence} and \citet{barbera2010individual} for social choice setting, \citet{barbera2016group} for general private good economies, and \citet{mandal2023equivalence} for two-sided matching markets.} In fact, we provide an example (Example \ref{example SP but not WGSP}) to demonstrate that not every stable and strategy-proof matching rule on such a domain is weakly group strategy-proof.
	
	Our results highlight the difficulty of achieving stability while maintaining other incentive properties, even when we only consider single-peaked preferences. Our analysis reveals that a strong domain restriction is required for stability and strategy-proofness to coexist on (tree-)single-peaked subdomains, and that stability and non-bossiness are incompatible on such domains.

	\subsection{Related literature}

	\citet{alcalde1994top} is the first paper to explore the possibility of coexistence between stability and strategy-proofness by restricting the domain. They show that the domain satisfying top dominance for women is sufficient for the \textit{men-proposing DA (MPDA) rule} to be strategy-proof, regardless of the domain being rich, anonymous, or a (tree-)single-peaked subdomain. Later, \citet{mandal2023equivalence} extends their finding by proving that the same condition is sufficient for the MPDA rule to be weakly group strategy-proof. \citeauthor{mandal2023equivalence}'s result plays a role in proving our characterization result (Theorem$^*$ \ref{theo sp-wgsp tree single peaked}).
	
	\citet{alcalde1994top} also identify top dominance as a necessary condition. They show that when men's preferences are unrestricted, top dominance for women is necessary for the existence of a stable and strategy-proof matching rule. While our characterization result resembles \citeauthor{alcalde1994top}'s, we differ by identifying top dominance as a necessary condition for all possible (tree-)single-peaked subdomains, whereas their result is based on the assumption that one side of the market has unrestricted preferences.
	
	Another paper related to our findings is \citet{kojima2010impossibility}, which shows the incompatibility between stability and non-bossiness when the agents' preferences are unrestricted. Our impossibility result (Proposition$^*$ \ref{proposition no stable and nb rule for atleast 5 tree}) is much stronger. It shows that the coexistence of stability and non-bossiness is impossible on any (tree-)single-peaked subdomain, provided either side of the market has at least five agents.
	
	One small but important difference between the setting considered by \citet{alcalde1994top}, \citet{kojima2010impossibility}, and \citet{mandal2023equivalence} and our setting, apart from having a balanced market, is how the ``choice of remaining unmatched'' (henceforth, the \textit{outside option}) is treated. In the aforementioned papers, the outside option can be freely ranked by agents, while in our setting, it is ranked last by default. Due to this difference, one has to be cautious when comparing results for these settings.

	\subsection{Organization of the paper}

	The paper is structured as follows: Section \ref{section preliminaries} introduces basic concepts and notations that we use throughout the paper. We describe our model, define matching rules, and discuss their standard properties. Section \ref{section single-peakedness} introduces the notion of single-peakedness. In Section \ref{section DA}, we present the DA algorithm. Section \ref{section results} presents the results. Section \ref{section tree single peaked} generalizes our results to the single-peaked subdomains on trees. The appendix contains the proofs.

	\section{Preliminaries}\label{section preliminaries}

	\subsection{Basic notions and notations}

	For a finite set $X$, let $\mathbb{L}(X)$ denote the set of all \textit{strict linear order}s over $X$.\footnote{A \textbf{\textit{strict linear order}} is a complete, asymmetric, and transitive binary relation.} An element of $\mathbb{L}(X)$ is called a \textbf{\textit{preference}} over $X$. 
	For a preference $P \in \mathbb{L}(X)$ and distinct $x, y \in X$, $x \mathrel{P} y$ is interpreted as ``$x$ is preferred to $y$ according to $P$''. 
	For $P \in \mathbb{L}(X)$, let $R$ denote the weak part of $P$, i.e., for all $x, y \in X$, $x \mathrel{R} y$ if and only if \big[$x \mathrel{P} y$ or $x=y$\big]. 
	Finally, for $P \in \mathbb{L}(X)$, let $\tau(P)$ denote the most preferred element in $X$ according to $P$, i.e., $\tau(P) = x$ if and only if \big[$x \in X$ and $x \mathrel{P} y$ for all $y \in X \setminus \{x\}$\big].

	\subsection{Model}
	
	There are two finite disjoint sets of agents, the set of \textit{men} $M = \{m_1, \ldots, m_n\}$, and the set of equally many \textit{women} $W = \{w_1, \ldots, w_n\}$. Let $A = M \cup W$ be the set of all agents. To avoid trivialities, we assume $n \geq 3$ throughout the paper.
	
	Each man $m$ has a preference $P_m$ over $W$ and each woman $w$ has a preference $P_w$ over $M$. We denote by $\mathcal{P}_a$ the \textit{set of admissible preferences} for agent $a$. Clearly, $\mathcal{P}_m \subseteq \mathbb{L}(W)$ for all $m \in M$ and $\mathcal{P}_w \subseteq \mathbb{L}(M)$ for all $w \in W$. A \textbf{\textit{preference profile}}, denoted by $P_A = (P_{m_1}, \ldots, P_{m_n}, P_{w_1}, \ldots, P_{w_n})$, is an element of the Cartesian product $\mathcal{P}_A := \underset{i=1}{\overset{n}{\prod}}\mathcal{P}_{m_i} \times \underset{j=1}{\overset{n}{\prod}}\mathcal{P}_{w_j}$, which represents a collection of preferences -- one for each agent. 
	We call $\mathcal{P}_A$ the \textit{domain of preference profiles}, or simply the \textit{domain}. 
	Furthermore, as is the convention, $P_{-a}$ denotes a collection of preferences of all agents except for $a$. Also, for $A' \subseteq A$, let $P_{A'}$ denote a collection of preferences of all agents in $A'$, and let $P_{-A'}$ denote a collection of preferences of all agents not in $A'$.
	
	We now introduce two standard properties of domains: \textit{richness} and \textit{anonymity}.
	
	The first property, \textit{richness}, reflects the idea that every agent should be allowed to express a strong preference for any potential match, ensuring flexibility in individual preferences.

	\begin{definition}[Richness]
		The domain of preference profiles $\mathcal{P}_A$ is \textit{\textbf{rich}} if for every $m \in M$ and every $w \in W$,
		\begin{enumerate}[(i)]
			\item there exists $P_m \in \mathcal{P}_m$ such that $\tau(P_m) = w$, and
			
			\item there exists $P_w \in \mathcal{P}_w$ such that $\tau(P_w) = m$.
		\end{enumerate}
	\end{definition}

	The second property, \textit{anonymity}, is based on the idea of a common-domain approach. Anonymity requires every man to have the same set of admissible preferences, and it requires the same for women.

	\begin{definition}[Anonymity]\label{definition anonymity}
		The domain of preference profiles $\mathcal{P}_A$ is \textbf{\textit{anonymous}} if 
		\begin{enumerate}[(i)]
			\item $\mathcal{P}_{m} = \mathcal{P}_{\tilde{m}}$ for all $m, \tilde{m} \in M$, and
			
			\item $\mathcal{P}_{w} = \mathcal{P}_{\tilde{w}}$ for all $w, \tilde{w} \in W$.
		\end{enumerate}
	\end{definition}

	Whenever $\mathcal{P}_A$ is an anonymous domain, for ease of presentation, we denote the common sets of admissible preferences for men and women by $\mathcal{P}_{men}$ and $\mathcal{P}_{women}$, respectively.

	\subsection{Matching rules and their stability}

	A \textbf{\textit{matching}} (between $M$ and $W$) is a one-to-one function $\mu : A \to A$ such that 
	\begin{enumerate}[(i)]
		\item\label{item matching definition 1} $\mu(m) \in W$ for all $m \in M$,
		
		\item\label{item matching definition 2} $\mu(w) \in M$ for all $w \in W$, and
		
		\item\label{item matching definition 3} $\mu(m) = w$ if and only if $\mu(w) = m$ for all $m \in M$ and all $w \in W$.\footnote{Notice that Condition \ref{item matching definition 2} is redundant in our framework. Since $|M| = |W|$ and $\mu$ is a one-to-one function, Condition \ref{item matching definition 2} follows from Condition \ref{item matching definition 1}.}
	\end{enumerate}
	Here, $\mu(m) = w$ means man $m$ and woman $w$ are matched to each other under the matching $\mu$. We denote the set of all matchings by $\mathcal{M}$.
	
	We now introduce the well-known notion of \textit{stability} \citep{gale1962college}, which is a desirable property for a matching.

	\begin{definition}[Stability]
		A matching $\mu$ is \textbf{\textit{stable}} at a preference profile $P_A \in \mathcal{P}_A$ if there exists no man-woman pair $(m,w) \in M \times W$ such that $w \mathrel{P_m} \mu(m)$ and $m \mathrel{P_w} \mu(w)$.
	\end{definition}

	We denote by $\mathcal{C}(P_A)$ the set of all stable matchings at a preference profile $P_A$.
	
	A \textbf{\textit{matching rule}} is a function $\varphi: \mathcal{P}_A \to \mathcal{M}$. For a matching rule $\varphi: \mathcal{P}_A \to \mathcal{M}$ and a preference profile $P_A \in \mathcal{P}_A$, let $\varphi_a(P_A)$ denote the match of agent $a$ by $\varphi$ at $P_A$. 
	
	A matching rule $\varphi: \mathcal{P}_A \to \mathcal{M}$ is \textbf{\textit{stable}} if for every $P_A \in \mathcal{P}_A$, $\varphi(P_A)$ is stable at $P_A$.

	\subsection{Incentive properties of matching rules}

	In practice, matching rules are often designed to satisfy incentive properties. Three well-studied such requirements are \textit{strategy-proofness}, \textit{weak group strategy-proofness}, and \textit{strong group strategy-proofness}.

	\begin{definition}[Incentive properties]\label{definition incentive properties}
		A matching rule $\varphi: \mathcal{P}_A \to \mathcal{M}$ is
		\begin{enumerate}[(i)]
			\item \textbf{\textit{strategy-proof}} if for every $P_A \in \mathcal{P}_A$, every $a \in A$, and every $\tilde{P}_a \in \mathcal{P}_a$, we have $\varphi_a(P_A) \mathrel{R_a} \varphi_a(\tilde{P}_a, P_{-a})$.
			
			\item\label{item wgsp} \textit{\textbf{weakly group strategy-proof}} if for every $P_A \in \mathcal{P}_A$, there do not exist a set of agents $\tilde{A} \subseteq A$ and a preference profile $\tilde{P}_{\tilde{A}} \in \mathcal{P}_{\tilde{A}}$ such that $\varphi_a(\tilde{P}_{\tilde{A}},P_{-\tilde{A}}) \mathrel{P_a} \varphi_a(P_A)$ for all $a \in \tilde{A}$.
			
			\item\label{item gsp} \textit{\textbf{strongly group strategy-proof}} if for every $P_A \in \mathcal{P}_A$, there do not exist a set of agents $\tilde{A} \subseteq A$ and a preference profile $\tilde{P}_{\tilde{A}} \in \mathcal{P}_{\tilde{A}}$ such that
			\begin{enumerate}[(a)]
				\item $\varphi_a(\tilde{P}_{\tilde{A}},P_{-\tilde{A}}) \mathrel{R_a} \varphi_a(P_A)$ for all $a \in \tilde{A}$, and
				
				\item $\varphi_b(\tilde{P}_{\tilde{A}},P_{-\tilde{A}}) \mathrel{P_b} \varphi_b(P_A)$ for at least one $b \in \tilde{A}$.
			\end{enumerate}
		\end{enumerate}
	\end{definition}

	Note that in Definition \ref{definition incentive properties}, $\tilde{A}$ may consist of both men and women in \ref{item wgsp} and \ref{item gsp}.
	
	By definition, strong group strategy-proofness is stronger than weak group strategy-proofness, and both are stronger than strategy-proofness. If a matching rule $\varphi$ on $\mathcal{P}_A$ is not strategy-proof, then there exist a preference profile $P_A \in \mathcal{P}_A$, an agent $a \in A$, and a preference $\tilde{P}_a \in \mathcal{P}_a$ of agent $a$ such that $\varphi_a(\tilde{P}_a, P_{-a}) \mathrel{P_a} \varphi_a(P_A)$. In such cases, we say that \textit{$\varphi$ is manipulable at $P_A$ by agent $a$ via $\tilde{P}_a$}.

	\section{Single-peakedness: A natural restriction on preferences}\label{section single-peakedness}

	In many real-world scenarios, agents' preferences may have specific restricted structures. A natural restriction is as follows: the agents on each side are (linearly) ordered based on specific criteria (potentially different criteria for each side), and the preferences respect these orderings in the sense that as an agent moves away from their most preferred choice, in either direction (according to the ordering of the agents on the other side), their welfare declines. Such restriction is known as \textit{single-peakedness} \citep{black1948rationale} in the literature, which we formally present below for our setting.
	
	Let $\prec_M$ be a prior (linear) ordering over $M$ and $\prec_W$ a prior (linear) ordering over $W$. 
	A preference $P \in \mathbb{L}(W)$ is \textit{\textbf{single-peaked} with respect to $\prec_W$} if for every $w, \tilde{w} \in W$, $\big[\tau(P) \preceq_W w \prec_W \tilde{w} \mbox{ or } \tilde{w} \prec_W w \preceq_W \tau(P)\big]$ implies $w \mathrel{P} \tilde{w}$.\footnote{$\preceq_W$ denotes the weak part of $\prec_W$.} Similarly, a preference $P \in \mathbb{L}(M)$ is \textit{\textbf{single-peaked} with respect to $\prec_M$} if for every $m, \tilde{m} \in M$, $\big[\tau(P) \preceq_M m \prec_M \tilde{m} \mbox{ or } \tilde{m} \prec_M m \preceq_M \tau(P)\big]$ implies $m \mathrel{P} \tilde{m}$. 
	Let $\mathbb{S}(\prec_M)$ and $\mathbb{S}(\prec_W)$ denote the sets of all single-peaked preferences with respect to $\prec_M$ and $\prec_W$, respectively. 
	
	Throughout this paper, whenever we write that $\mathcal{P}_A$ is the \textit{maximal single-peaked domain}, we mean $\mathcal{P}_m = \mathbb{S}(\prec_W)$ for all $m \in M$ and $\mathcal{P}_w = \mathbb{S}(\prec_M)$ for all $w \in W$. Also, whenever we write that $\mathcal{P}_A$ is a \textit{single-peaked subdomain}, we mean $\mathcal{P}_m \subseteq \mathbb{S}(\prec_W)$ for all $m \in M$ and $\mathcal{P}_w \subseteq \mathbb{S}(\prec_M)$ for all $w \in W$.

	\section{Deferred acceptance: A stable matching rule}\label{section DA}

	\textit{Deferred acceptance (DA) algorithm} \citep{gale1962college} is the salient algorithm in two-sided matching markets for its theoretical appeal. This algorithm also plays an instrumental role in our results. Hence, we briefly describe this algorithm for completeness' sake.
	
	There are two types of the DA algorithm: the \textit{men-proposing DA (MPDA) algorithm} and the \textit{women-proposing DA (WPDA) algorithm}. In the following, we provide a description of the MPDA algorithm at a preference profile $P_A$. The same of the WPDA algorithm can be obtained by interchanging the roles of men and women in the MPDA algorithm.

	\begin{itemize}[leftmargin = 1.35cm]
		\item[\textbf{\textit{Step 1.}}] Each man $m$ proposes to his most preferred woman (according to $P_m$). Every woman $w$, who has at least one proposal, tentatively keeps her most preferred man (according to $P_w$) among these proposals and rejects the rest.
		
		\item[\textbf{\textit{Step $2$.}}] Every man $m$ who was rejected at the previous step proposes to his next preferred woman. Every woman $w$ who has at least one proposal, including any proposal tentatively kept from the earlier steps, tentatively keeps her most preferred man among these proposals and rejects the rest.
	\end{itemize}

	This procedure is then repeated from Step 2 till a step such that each woman has a proposal. At this step, the proposals tentatively accepted by women become permanent. This completes the description of the MPDA algorithm.\medskip
	
	Let $D^M$ and $D^W$ denote the matching rules associated with the MPDA and WPDA algorithms, respectively. In other words, the matching rules $D^M$ and $D^W$ select, at each preference profile $P_A$, the matchings $D^M(P_A)$ and $D^W(P_A)$, the outcomes of the MPDA and WPDA algorithms at $P_A$, respectively.
	
	The following remark summarizes some properties of the matching rules $D^M$ and $D^W$, which we will use in the proofs (see \citet{gale1962college} and \citet{mcvitie1971stable} for details of these properties). These properties hold regardless of whether the domain is rich, anonymous, or a single-peaked subdomain.

	\begin{remark}\label{remark properties of DA}
		Let $\mathcal{P}_A$ be an arbitrary domain of preference profiles. For every $P_A \in \mathcal{P}_A$,
		\begin{enumerate}[(i)]
			\item $D^M(P_A), D^W(P_A) \in \mathcal{C}(P_A)$, and
			
			\item for every $\mu \in \mathcal{C}(P_A)$,
			\begin{enumerate}[(a)]
				\item $D^M_m(P_A) \mathrel{R_m} \mu(m) \mathrel{R_m} D^W_m(P_A)$ for all $m \in M$, and
				
				\item $D^W_w(P_A) \mathrel{R_w} \mu(w) \mathrel{R_w} D^M_w(P_A)$ for all $w \in W$.
			\end{enumerate}
		\end{enumerate}
	\end{remark}

	\section{Results: Characterizations and impossibilities}\label{section results}

	In this section, we investigate the potential existence of stable and (weakly/strongly group) strategy-proof matching rules on single-peaked subdomains. For this purpose, we first introduce a domain restriction, called \textit{top dominance} \citep{alcalde1994top}.\footnote{\citet{alcalde1994top} introduce top dominance in a setting with outside options. We have reformulated this domain restriction for our setting (i.e., without outside options).} 
	We use the following notation to present it: for distinct $m, \tilde{m}, m' \in M$ and $w \in W$, whenever we write $(\cdot m \cdot \tilde{m} \cdot m' \cdot) \in \mathcal{P}_w$, we mean that there exists a preference $P \in \mathcal{P}_w$ such that $m \mathrel{P} \tilde{m} \mathrel{P} m'$. Here, $m$ and $\tilde{m}$ need not be consecutively ranked in $P$, nor do $\tilde{m}$ and $m'$. Furthermore, $m$ need not be the most preferred man according to $P$. Clearly, $(\cdot m \cdot \tilde{m} \cdot m' \cdot) \notin \mathcal{P}_w$ is interpreted as ``there exists no preference $P \in \mathcal{P}_w$ such that $m \mathrel{P} \tilde{m} \mathrel{P} m'$''.

	\begin{definition}[Top dominance]
		The domain of preference profiles $\mathcal{P}_A$ satisfies \textit{\textbf{top dominance} for women} if for every $w \in W$, $\mathcal{P}_w$ satisfies the following condition: for every $m, \tilde{m}, m' \in M$,
		\begin{equation*}
			(\cdot m \cdot \tilde{m} \cdot m' \cdot) \in \mathcal{P}_w \implies (\cdot m \cdot m' \cdot \tilde{m} \cdot) \notin \mathcal{P}_w.
		\end{equation*}
	\end{definition}

	We define \textit{\textbf{top dominance} for men} in a similar way.
	
	Note that whenever $\mathcal{P}_A$ satisfies top dominance for women, no two distinct admissible preferences for a woman can have the same most preferred man. The top choice determines the rest of the preference, hence the name for the domain restriction. 
	An instance of a single-peaked subdomain satisfying top dominance for women is when the sets of admissible preferences for women consist of \textit{left-biased single-peaked preferences} \citep{alva2017manipulation}.\footnote{A single-peaked preference $P \in \mathbb{S}(\prec_M)$ is \textit{\textbf{left-biased}} if for every $m, \tilde{m} \in M \setminus \{\tau(P)\}$, $m \prec_M \tau(P) \prec_M \tilde{m}$ implies $m \mathrel{P} \tilde{m}$. \citet{klaus2013relation} refer to these preferences as ``left-right single-peaked'' and provide economic examples where they occur.}
	
	We now present the first main result of this paper. It characterizes all rich and anonymous single-peaked subdomains on which a stable and (weakly group) strategy-proof matching rule exists. It further demonstrates that on such domains, at least one of the matching rules $D^M$ and $D^W$ is stable and weakly group strategy-proof.

	\begin{theorem}\label{theo sp-wgsp classical single peaked}
		Suppose $\mathcal{P}_A$ is a rich and anonymous single-peaked subdomain. The following statements are equivalent:
		\begin{enumerate}[(a)]
			\item $\mathcal{P}_A$ satisfies top dominance for at least one side of the market.
			
			\item There exists a stable and strategy-proof matching rule on $\mathcal{P}_A$.
			
			\item There exists a stable and weakly group strategy-proof matching rule on $\mathcal{P}_A$.
			
			\item\label{item DA is wgsp classical} At least one of the matching rules $D^M$ and $D^W$ satisfies stability and weak group strategy-proofness.
		\end{enumerate}
	\end{theorem}

	Later, in Section \ref{section tree single peaked}, we prove a stronger result (Theorem$^*$ \ref{theo sp-wgsp tree single peaked}), and therefore, we skip the proof of Theorem \ref{theo sp-wgsp classical single peaked}.
	
	There are two important points to note about Theorem \ref{theo sp-wgsp classical single peaked}. First, we cannot replace ``at least one of the matching rules $D^M$ and $D^W$'' with ``both of the matching rules $D^M$ and $D^W$'' in \ref{item DA is wgsp classical}. To illustrate this, consider a rich and anonymous single-peaked subdomain that satisfies top dominance for women but not for men. In this case, the matching rule $D^M$ is stable and weakly group strategy-proof (see Theorem \ref{theorem mandal result}), but the matching rule $D^W$ may not even be strategy-proof. Second, while Theorem \ref{theo sp-wgsp classical single peaked} establishes that under richness and anonymity, the single-peaked subdomains on which stability and strategy-proofness are compatible are the same ones on which stability and weak group strategy-proofness are compatible, it does not imply that every stable and strategy-proof matching rule on such a domain is weakly group strategy-proof. In fact, we provide an example (Example \ref{example SP but not WGSP}) in Appendix \ref{appendix example} to demonstrate that not every stable and strategy-proof matching rule on such a domain is weakly group strategy-proof. (Example \ref{example SP but not WGSP} also highlights an important observation: not every stable and strategy-proof matching rule is simply a combination of the matching rules $D^M$ and $D^W$. A detailed explanation is provided in Appendix \ref{appendix example}.)
	
	Since satisfying top dominance (for at least one side of the market) is both a necessary and sufficient condition for a rich and anonymous single-peaked subdomain to have stable and (weakly group) strategy-proof matching rules (as demonstrated by Theorem \ref{theo sp-wgsp classical single peaked}), it is natural to ask what types of domain structures emerge within the single-peaked subdomains under top dominance. One example, as mentioned earlier, is when the sets of admissible preferences consist of left-biased single-peaked preferences. In Appendix \ref{appendix domain structure}, we provide a detailed discussion with additional examples of such domains.
	
	Our next result is an impossibility result, which shows the incompatibility between stability and strategy-proofness on the maximal single-peaked domain. To see this, notice that the maximal single-peaked domain does not satisfy top dominance for either side of the market. Because of this, and since the maximal single-peaked domain is rich and anonymous, our result follows from Theorem \ref{theo sp-wgsp classical single peaked}.

	\begin{corollary}\label{corollary impossibility on maximal single peaked}
		On the maximal single-peaked domain $\mathbb{S}^n(\prec_W) \times \mathbb{S}^n(\prec_M)$, no stable matching rule is strategy-proof.
	\end{corollary}

	We now focus on the compatibility between stability and strong group strategy-proofness on single-peaked subdomains. We begin by exploring the compatibility between stability and \textit{non-bossiness} -- a weaker concept than strong group strategy-proofness. Non-bossiness is a classic normative property, which requires that no agent can change others' matches without changing their own.

	\begin{definition}[Non-bossiness]
		A matching rule $\varphi : \mathcal{P}_A \to \mathcal{M}$ is \textbf{\textit{non-bossy}} if for every $P_A \in \mathcal{P}_A$, every $a \in A$, and every $\tilde{P}_a \in \mathcal{P}_a$,
		\begin{equation*}
			\varphi_a(P_A) = \varphi_a(\tilde{P}_a, P_{-a}) \implies \varphi(P_A) = \varphi(\tilde{P}_a, P_{-a}).
		\end{equation*}
	\end{definition}

	Our next result demonstrates that there is an incompatibility between stability and non-bossiness on single-peaked subdomains. Specifically, it shows that whenever either side of the market has at least five agents, there are no rich and anonymous single-peaked subdomains with stable and non-bossy matching rules.

	\begin{proposition}\label{proposition no stable and nb rule for atleast 5}
		Suppose $\mathcal{P}_A$ is a rich and anonymous single-peaked subdomain. If $n \geq 5$, then no stable matching rule on $\mathcal{P}_A$ is non-bossy.
	\end{proposition}

	We skip the proof of Proposition \ref{proposition no stable and nb rule for atleast 5} as we prove a stronger result (Proposition$^*$ \ref{proposition no stable and nb rule for atleast 5 tree}) in Section \ref{section tree single peaked}.
	
	We now present our next main result, which shows the incompatibility between stability and strong group strategy-proofness on rich and anonymous single-peaked subdomains. This incompatibility result follows from Proposition \ref{proposition no stable and nb rule for atleast 5}, as strong group strategy-proofness implies non-bossiness.

	\begin{theorem}\label{theorem no stable and gsp}
		Suppose $\mathcal{P}_A$ is a rich and anonymous single-peaked subdomain. If $n \geq 5$, then no stable matching rule on $\mathcal{P}_A$ is strongly group strategy-proof.
	\end{theorem}

	\begin{note}
		Proposition \ref{proposition no stable and nb rule for atleast 5} and Theorem \ref{theorem no stable and gsp} do not hold when $n \in \{3, 4\}$. In these cases, there exist rich and anonymous single-peaked subdomains that admit stable and strongly group strategy-proof (consequently, non-bossy) matching rules.\footnote{However, neither of the matching rules $D^M$ and $D^W$, nor any combination of them, satisfies strong group strategy-proofness on these domains.}
	\end{note}

	\section{Generalization: Single-peakedness on trees}\label{section tree single peaked}

	So far, we have focused on \textit{classical single-peaked preferences}, where single-peakedness is defined with respect to a \textit{linear} prior ordering over the alternatives. However, many real-world decision-making scenarios are more intricate: alternatives are often interconnected as nodes within a network, rather than arranged along a simple linear spectrum. In these cases, agents' preferences are naturally shaped by the structure of the underlying network, highlighting the need to move from classical single-peakedness to a network-based notion of single-peakedness.
	
	For example, imagine a scenario where all the agents are located on a ``road network'', and their welfare diminishes with increasing distance. In such cases, a man will prefer to be matched with a woman over another woman if the first woman is located on every possible road between the man and the other woman. This illustrates how the network influences preference orderings, requiring us to account for topological constraints in our setting. In this paper, we focus on cases where the network exhibits a \textit{tree} structure, which simplifies the analysis while many desirable properties of the linear model are retained.
	
	In this section, we introduce the concept of \textit{single-peakedness on trees} \citep{demange1982single}, and explore how our previous results can be extended within this framework.

	\subsection{Single-peakedness on trees}

	We begin by presenting a few notions and notations related to graph theory that we use throughout this section. A \textit{path} in a graph is a sequence of distinct nodes such that every two consecutive nodes form an edge. A \textbf{\textit{tree}}, denoted by $T$, is an undirected graph in which every two nodes are connected by exactly one path. For a tree $T$, we denote its set of nodes by $V(T)$.
	
	Let $T_M$ be a tree with $M$ as its set of nodes, and let $T_W$ be a tree with $W$ as its set of nodes. 
	A preference $P \in \mathbb{L}(W)$ is \textit{\textbf{single-peaked} with respect to $T_W$} if for every distinct $w, \tilde{w} \in W$ with $w$ being on the (unique) path between $\tau(P)$ and $\tilde{w}$ in $T_W$, we have $w \mathrel{P} \tilde{w}$. Similarly, a preference $P \in \mathbb{L}(M)$ is \textit{\textbf{single-peaked} with respect to $T_M$} if for every distinct $m, \tilde{m} \in M$ with $m$ being on the (unique) path between $\tau(P)$ and $\tilde{m}$ in $T_M$, we have $m \mathrel{P} \tilde{m}$. 
	Let $\mathbb{S}(T_M)$ and $\mathbb{S}(T_W)$ denote the sets of all single-peaked preferences with respect to $T_M$ and $T_W$, respectively.
	
	Classical single-peakedness is a special case of single-peakedness on trees. Example \ref{example classical single-peaked is a special case} below illustrates this.

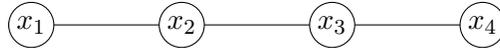
\begin{figure}[H] 
		\centering
		\begin{tikzpicture}[->]
			\begin{scope}
				\draw[draw = black] (1,1) circle (.3) (3,1) circle (.3) (5,1) circle (.3) (7,1) circle (.3) (9,1) circle (.3);
				\node at (1,1) {$m_1$}; \node at (3,1) {$m_2$}; \node at (5,1) {$m_3$}; \node at (7,1) {$m_4$}; \node at (9,1) {$m_5$};
				\path (1.3,1) edge[-] node {} (2.7,1);
				\path (3.3,1) edge[-] node {} (4.7,1);
				\path (5.3,1) edge[-] node {} (6.7,1);
				\path (7.3,1) edge[-] node {} (8.7,1);
			\end{scope}
		\end{tikzpicture}
		\caption{Linear tree $T_M$ for Example \ref{example classical single-peaked is a special case}}
		\label{figure tree for classical single-peaked}
	\end{figure}

	\begin{example}[Classical single-peakedness is a special case of single-peakedness on trees]\label{example classical single-peaked is a special case}
		Suppose $M = \{m_1, m_2, m_3, m_4, m_5\}$ with a prior (linear) ordering $\prec_M$ such that $m_1 \prec_M m_2 \prec_M m_3 \prec_M m_4 \prec_M m_5$. It is straightforward to verify that $\mathbb{S}(\prec_M) = \mathbb{S}(T_M)$, where $T_M$ is the linear tree presented in Figure \ref{figure tree for classical single-peaked}.
		\hfill
		$\Diamond$
	\end{example}

	It is natural to wonder about the structure of single-peaked preferences with respect to a non-linear tree; below is such an example.

	\begin{example}[Single-peaked preferences with respect to a non-linear tree]\label{example starfish shaped}
		Suppose $M = \{m_1, m_2, m_3, m_4, m_5\}$. Consider the tree $T_M$ presented in Figure \ref{figure tree for starfish}. Notice that $T_M$ has a \textit{star}-structure.\footnote{A \textbf{\textit{star}}, denoted by $S_k$, is a tree with 
			\begin{enumerate}[(i)]
				\item one internal node and $k$ leaves (terminal nodes) when $k > 1$, or
				\item no internal nodes and $k + 1$ leaves when $k \leq 1$.
			\end{enumerate}
			See \href{https://en.wikipedia.org/wiki/Star_(graph_theory)}{Star (graph theory)} for details.}
		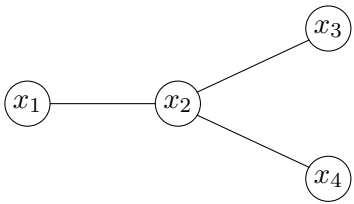
\begin{figure}[H] 
			\centering
			\begin{tikzpicture}[->]
				\begin{scope}
					\draw[draw = black] (1,2) circle (.3) (1,0) circle (.3) (3,1) circle (.3) (5,2) circle (.3) (5,0) circle (.3);
					\node at (1,2) {$m_2$}; \node at (1,0) {$m_3$}; \node at (3,1) {$m_1$}; \node at (5,2) {$m_4$}; \node at (5,0) {$m_5$};
					\path (2.75,1.15) edge[-] node {} (1.25,1.85);
					\path (2.75,.85) edge[-] node {} (1.25,.15);
					\path (3.25,1.15) edge[-] node {} (4.75,1.85);
					\path (3.25,.85) edge[-] node {} (4.75,.15);
				\end{scope}
			\end{tikzpicture}
			\caption{Tree $T_M$ for Example \ref{example starfish shaped}}
			\label{figure tree for starfish}
		\end{figure}
		
		In this case, $\mathbb{S}(T_M)$ consists of all the preferences that rank $m_1$ first or second. We leave the verification of this fact to the reader. 
		\hfill
		$\Diamond$
	\end{example}

	Examples \ref{example classical single-peaked is a special case} and \ref{example starfish shaped} together illustrate an important observation: the number of single-peaked preferences depends on the structure of the underlying tree. For instance, in Example \ref{example classical single-peaked is a special case}, $|\mathbb{S}(T_M)| = 16$, whereas in Example \ref{example starfish shaped}, $|\mathbb{S}(T_M)| = 48$. Specifically, the number of single-peaked preferences in $\mathbb{S}(T_M)$ ranges from $2^{n-1}$ (when $T_M$ is linear) to $2(n-1)!$ (when $T_M$ is a star), where $|M| = n$.
	
	Throughout this section, whenever we write that $\mathcal{P}_A$ is the \textit{maximal tree-single-peaked domain}, we mean $\mathcal{P}_m = \mathbb{S}(T_W)$ for all $m \in M$ and $\mathcal{P}_w = \mathbb{S}(T_M)$ for all $w \in W$. Also, whenever we write that $\mathcal{P}_A$ is a \textit{tree-single-peaked subdomain}, we mean $\mathcal{P}_m \subseteq \mathbb{S}(T_W)$ for all $m \in M$ and $\mathcal{P}_w \subseteq \mathbb{S}(T_M)$ for all $w \in W$.

	\subsection{Generalized results}

	All of our findings for classical single-peaked subdomains (in Section \ref{section results}) can be generalized to tree-single-peaked subdomains; we present these generalizations in this subsection.
	
	Our first result in this section is the generalized version of Theorem \ref{theo sp-wgsp classical single peaked}. It characterizes all rich and anonymous tree-single-peaked subdomains on which a stable and (weakly group) strategy-proof matching rule exists. It further demonstrates that on such domains, at least one of the matching rules $D^M$ and $D^W$ is stable and weakly group strategy-proof.

	\begin{theorem*}\label{theo sp-wgsp tree single peaked}
		Suppose $\mathcal{P}_A$ is a rich and anonymous tree-single-peaked subdomain. The following statements are equivalent:
		\begin{enumerate}[(a)]
			\item\label{item TD} $\mathcal{P}_A$ satisfies top dominance for at least one side of the market.
			
			\item\label{item stable and sp} There exists a stable and strategy-proof matching rule on $\mathcal{P}_A$.
			
			\item\label{item stable and wgsp} There exists a stable and weakly group strategy-proof matching rule on $\mathcal{P}_A$.
			
			\item\label{item DA is wgsp} At least one of the matching rules $D^M$ and $D^W$ satisfies stability and weak group strategy-proofness.
		\end{enumerate}
	\end{theorem*}

	The proof of Theorem$^*$ \ref{theo sp-wgsp tree single peaked} is relegated to Appendix \ref{appendix proof of theorem sp-wgsp tree single peaked}; here, we provide an outline of it. The implication that \ref{item TD} implies \ref{item DA is wgsp} in Theorem$^*$ \ref{theo sp-wgsp tree single peaked} follows from a result by \citet{mandal2023equivalence}, which establishes that the matching rule $D^M$ becomes weakly group strategy-proof when the domain satisfies top dominance for women (see Theorem \ref{theorem mandal result}). The implications that \ref{item DA is wgsp} implies \ref{item stable and wgsp} and \ref{item stable and wgsp} implies \ref{item stable and sp} are straightforward. These three implications hold regardless of the domain being rich, anonymous, or a tree-single-peaked subdomain. To prove that \ref{item stable and sp} implies \ref{item TD}, we demonstrate the contraposition. In cases where the domain $\mathcal{P}_A$ does not satisfy top dominance for either side of the market, we construct a preference profile (in $\mathcal{P}_A$) based on richness, anonymity, and tree-single-peakedness, and then show that every stable matching rule on $\mathcal{P}_A$ is manipulable at the constructed preference profile by some agent.
	
	Our next result -- the generalized version of Corollary \ref{corollary impossibility on maximal single peaked} -- is an impossibility result. It demonstrates the incompatibility between stability and strategy-proofness on the maximal tree-single-peaked domain, regardless of the structures of the underlying trees $T_M$ and $T_W$. To see this, notice that the maximal tree-single-peaked domain does not satisfy top dominance for either side of the market, regardless of the structures of $T_M$ and $T_W$. Because of this, and since the maximal tree-single-peaked domain is rich and anonymous, our result follows from Theorem$^*$ \ref{theo sp-wgsp tree single peaked}.

	\begin{corollary*}\label{corollary impossibility on maximal single peaked tree}
		On the maximal tree-single-peaked domain $\mathbb{S}^n(T_W) \times \mathbb{S}^n(T_M)$, no stable matching rule is strategy-proof.
	\end{corollary*}

	Our next two results are the generalized versions of Proposition \ref{proposition no stable and nb rule for atleast 5} and Theorem \ref{theorem no stable and gsp}. The first result shows that whenever either side of the market has at least five agents, there are no rich and anonymous tree-single-peaked subdomains with stable and non-bossy matching rules, regardless of the structures of $T_M$ and $T_W$. The second result demonstrates the incompatibility between stability and strong group strategy-proofness in a similar scenario. This second result follows from the first, as strong group strategy-proofness implies non-bossiness.

	\begin{proposition*}\label{proposition no stable and nb rule for atleast 5 tree}
		Suppose $\mathcal{P}_A$ is a rich and anonymous tree-single-peaked subdomain. If $n \geq 5$, then no stable matching rule on $\mathcal{P}_A$ is non-bossy.
	\end{proposition*}

	The proof of Proposition$^*$ \ref{proposition no stable and nb rule for atleast 5 tree} is relegated to Appendix \ref{appendix proof of proposition no stable and nb rule for atleast 5 tree}.

	\begin{theorem*}\label{theorem no stable and gsp tree}
		Suppose $\mathcal{P}_A$ is a rich and anonymous tree-single-peaked subdomain. If $n \geq 5$, then no stable matching rule on $\mathcal{P}_A$ is strongly group strategy-proof.
	\end{theorem*}

	\section{Concluding remarks}

	Our main contribution in this paper is identifying all (tree-)single-peaked subdomains on which stability and (weak/strong group) strategy-proofness are compatible, which we present through two key results. The first one characterizes all rich and anonymous (tree-)single-peaked subdomains with stable and (weakly group) strategy-proof matching rules. The second one is an impossibility result that shows the incompatibility between stability and strong group strategy-proofness on rich and anonymous (tree-)single-peaked subdomains.
	
	Our results highlight the difficulty of achieving stability while maintaining other incentive properties, even when we only consider single-peaked preferences. Our analysis reveals that a strong domain restriction is required for stability and strategy-proofness to coexist on rich and anonymous (tree-)single-peaked subdomains, and that stability and non-bossiness are incompatible on such domains.

	\appendixtocon
	\appendixtitletocon

	\renewcommand{\theequation}{\Alph{section}.\arabic{equation}}
	\renewcommand{\thetable}{\Alph{section}.\arabic{table}}
	\renewcommand{\thetheorem}{\Alph{section}.\arabic{theorem}}
	\renewcommand{\thedefinition}{\Alph{section}.\arabic{definition}}
	\renewcommand{\theobs}{\Alph{section}.\arabic{obs}}
	\renewcommand{\theremark}{\Alph{section}.\arabic{remark}}
	\renewcommand{\theexample}{\Alph{section}.\arabic{example}}

	\begin{appendices}

		\section{Proof of Theorem$^*$ \ref{theo sp-wgsp tree single peaked}}\label{appendix proof of theorem sp-wgsp tree single peaked}
		
		\setcounter{equation}{0}
		\setcounter{table}{0}
		\setcounter{theorem}{0}
		\setcounter{definition}{0}
		\setcounter{obs}{0}
		\setcounter{remark}{0}
		\setcounter{example}{0}

		We first prove a lemma that we use in the proof of Theorem$^*$ \ref{theo sp-wgsp tree single peaked}.

		\subsection{Lemma \ref{lemma TD on trees} and its proof}

		Before we discuss Lemma \ref{lemma TD on trees}, we make an observation: for distinct $m, \tilde{m}, m' \in M$, suppose $\tilde{m}$ is on the (unique) path between $m$ and $m'$ in $T_M$, then for any single-peaked preference with respect to $T_M$, $\tilde{m}$ cannot be the worst choice among $m$, $\tilde{m}$, and $m'$ (see \citet{demange1982single} for details). We can make a similar observation for $T_W$ as well. We present these observations formally as a remark for future reference.

		\begin{remark}\label{remark single-peaked on trees property}
			\begin{enumerate}[(i)]
				\item\label{item trees for men} For every distinct $m, \tilde{m}, m' \in M$ with $\tilde{m}$ being on the (unique) path between $m$ and $m'$ in $T_M$, we have $(\cdot m \cdot m' \cdot \tilde{m} \cdot) \notin \mathbb{S}(T_M)$ and $(\cdot m' \cdot m \cdot \tilde{m} \cdot) \notin \mathbb{S}(T_M)$.
				
				\item\label{item trees for women} For every distinct $w, \tilde{w}, w' \in W$ with $\tilde{w}$ being on the (unique) path between $w$ and $w'$ in $T_W$, we have $(\cdot w \cdot w' \cdot \tilde{w} \cdot) \notin \mathbb{S}(T_W)$ and $(\cdot w' \cdot w \cdot \tilde{w} \cdot) \notin \mathbb{S}(T_W)$.
			\end{enumerate}
		\end{remark}

		We now state and prove Lemma \ref{lemma TD on trees}.

		\begin{lemma}\label{lemma TD on trees}
			Suppose $\mathcal{P}_A$ is a rich and anonymous tree-single-peaked subdomain. 
			\begin{enumerate}[(a)]
				\item\label{item non-TD for women} If $\mathcal{P}_A$ does not satisfy top dominance for women, then there exist $m, \tilde{m}, m' \in M$ such that
				\begin{equation*}
					\big\{(\cdot m \cdot \tilde{m} \cdot m' \cdot), (\cdot m \cdot m' \cdot \tilde{m} \cdot), (\cdot \tilde{m} \cdot m \cdot m' \cdot) \big\} \in \mathcal{P}_{women}.
				\end{equation*}
				
				\item\label{item non-TD for men} If $\mathcal{P}_A$ does not satisfy top dominance for men, then there exist $w, \tilde{w}, w' \in W$ such that
				\begin{equation*}
					\big\{(\cdot w \cdot \tilde{w} \cdot w' \cdot), (\cdot w \cdot w' \cdot \tilde{w} \cdot), (\cdot \tilde{w} \cdot w \cdot w' \cdot) \big\} \in \mathcal{P}_{men}.
				\end{equation*}
			\end{enumerate}
		\end{lemma}

		\begin{proof}[\textbf{Proof of Lemma \ref{lemma TD on trees}}]
			We only provide the proof for part \ref{item non-TD for men}; the proof for part \ref{item non-TD for women} can be derived through similar reasoning.
			
			Since $\mathcal{P}_A$ is anonymous and does not satisfy top dominance for men, there exist distinct $w, \tilde{w}, w' \in W$ and two preferences $P_1, P_2 \in \mathcal{P}_{men}$ such that 
			\begin{subequations}\label{equation TD common}
				\begin{equation}\label{equation TD common 1}
					w \mathrel{P_1} \tilde{w} \mathrel{P_1} w', \mbox{ and}
				\end{equation}
				\begin{equation}\label{equation TD common 2}
					w \mathrel{P_2} w' \mathrel{P_2} \tilde{w}.
				\end{equation}
			\end{subequations}
			Moreover, since $\mathcal{P}_A$ is rich, there exists a preference $P_3 \in \mathcal{P}_{men}$ such that $\tau(P_3) = \tilde{w}$. Let $\tilde{\pi}$ denote the path between $w$ and $\tilde{w}$ in $T_W$, and $\pi'$ denote the path between $w$ and $w'$. We distinguish the following two cases.\medskip

			\noindent\textbf{\textsc{Case} 1}: Suppose $\tilde{\pi}$ and $\pi'$ have no common node other than $w$.
			
			Since $\tilde{\pi}$ and $\pi'$ have no common node other than $w$, the path between $\tilde{w}$ and $w'$ is the concatenation of $\tilde{\pi}$ and $\pi'$, and consequently, $w$ is on the path between $\tilde{w}$ and $w'$. This, along with the construction of $P_3$, implies that
			\begin{equation}\label{equation TD case 1}
				\tilde{w} \mathrel{P_3} w \mathrel{P_3} w'.
			\end{equation}
			\eqref{equation TD common} and \eqref{equation TD case 1} together complete the proof for Case 1.\medskip

			\noindent\textbf{\textsc{Case} 2}: Suppose $\tilde{\pi}$ and $\pi'$ have common nodes other than $w$. 
			
			On $\tilde{\pi}$, let $\hat{w}$ be the first node from $\tilde{w}$ such that $\hat{w}$ is on $\pi'$.

			\begin{claim}\label{claim distinct nodes}
				$w$, $\tilde{w}$, $w'$, and $\hat{w}$ are distinct.
			\end{claim}

			\begin{claimproof}[Proof of Claim \ref{claim distinct nodes}.]
				\begin{enumerate}[(i)]
					\item\label{item claim 1} Since $\tilde{\pi}$ and $\pi'$ have common nodes other than $w$, by the construction of $\hat{w}$, we have $\hat{w} \neq w$.
					
					\item\label{item claim 2} Suppose $\hat{w} = \tilde{w}$. Then, $\tilde{w}$ is on the path between $w$ and $w'$. This, along with Remark \ref{remark single-peaked on trees property}.\ref{item trees for women}, implies $(\cdot w \cdot w' \cdot \tilde{w} \cdot) \notin \mathbb{S}(T_W)$. However, since $\mathcal{P}_{men} \subseteq \mathbb{S}(T_W)$, this contradicts \eqref{equation TD common 2}.
					
					\item\label{item claim 3} Suppose $\hat{w} = w'$. Then, $w'$ is on the path between $\tilde{w}$ and $w$. This, along with Remark \ref{remark single-peaked on trees property}.\ref{item trees for women}, implies $(\cdot w \cdot \tilde{w} \cdot w' \cdot) \notin \mathbb{S}(T_W)$. However, since $\mathcal{P}_{men} \subseteq \mathbb{S}(T_W)$, this contradicts \eqref{equation TD common 1}.
				\end{enumerate}
				Since $w$, $\tilde{w}$, and $w'$ are distinct, \ref{item claim 1} -- \ref{item claim 3} together complete the proof of Claim \ref{claim distinct nodes}.
			\end{claimproof}

			By the construction of $\hat{w}$, it is on the path between $w$ and $\tilde{w}$. This, along with Claim \ref{claim distinct nodes} and Remark \ref{remark single-peaked on trees property}.\ref{item trees for women}, implies $(\cdot w \cdot \tilde{w} \cdot \hat{w} \cdot) \notin \mathbb{S}(T_W)$. Since $P_1 \in \mathcal{P}_{men} \subseteq \mathbb{S}(T_W)$, this, together with \eqref{equation TD common 1}, implies
			\begin{equation}\label{equation case 2 1}
				\hat{w} \mathrel{P_1} \tilde{w} \mathrel{P_1} w'.
			\end{equation}
			Similarly, by the construction of $\hat{w}$, it is on the path between $w$ and $w'$. This, along with Claim \ref{claim distinct nodes} and Remark \ref{remark single-peaked on trees property}.\ref{item trees for women}, implies $(\cdot w \cdot w' \cdot \hat{w} \cdot) \notin \mathbb{S}(T_W)$. Since $P_2 \in \mathcal{P}_{men} \subseteq \mathbb{S}(T_W)$, this, together with \eqref{equation TD common 2}, implies
			\begin{equation}\label{equation case 2 2}
				\hat{w} \mathrel{P_2} w' \mathrel{P_2} \tilde{w}.
			\end{equation}
			Furthermore, the path between $\tilde{w}$ and $w'$ is obtained by concatenating the paths $\tilde{\pi}_s$ and $\pi'_s$, where $\tilde{\pi}_s$ is the sub-path of $\tilde{\pi}$ between $\tilde{w}$ and $\hat{w}$, and $\pi'_s$ is the sub-path of $\pi'$ between $\hat{w}$ and $w'$, and consequently, node $\hat{w}$ is on the path between $\tilde{w}$ and $w'$. This, along with construction of $P_3$ and Claim \ref{claim distinct nodes}, implies that
			\begin{equation}\label{equation case 2 3}
				\tilde{w} \mathrel{P_3} \hat{w} \mathrel{P_3} w'.
			\end{equation}
			\eqref{equation case 2 1}, \eqref{equation case 2 2}, and \eqref{equation case 2 3} together complete the proof for Case 2.\medskip
			
			Since Cases 1 and 2 are exhaustive, this completes the proof of part \ref{item non-TD for men} of Lemma \ref{lemma TD on trees}.
		\end{proof}

		\subsection{Completion of the proof of Theorem$^*$ \ref{theo sp-wgsp tree single peaked}}

		To facilitate the proof of Theorem$^*$ \ref{theo sp-wgsp tree single peaked}, we present a result from \citet{mandal2023equivalence}, which holds regardless of the domain being rich, anonymous, or a (tree-)single-peaked subdomain.

		\begin{theorem}[\citealp{mandal2023equivalence}]\label{theorem mandal result}
			Let $\mathcal{P}_A$ be an arbitrary domain of preference profiles.
			\begin{enumerate}[(a)]
				\item $\mathcal{P}_A$ satisfying top dominance for women is a sufficient condition for the matching rule $D^M$ to be a stable and weakly group strategy-proof matching rule on $\mathcal{P}_A$.
				
				\item $\mathcal{P}_A$ satisfying top dominance for men is a sufficient condition for the matching rule $D^W$ to be a stable and weakly group strategy-proof matching rule on $\mathcal{P}_A$.
			\end{enumerate}
		\end{theorem}

		We now prove Theorem$^*$ \ref{theo sp-wgsp tree single peaked}.

		\begin{proof}[\textbf{Proof of Theorem$^*$ \ref{theo sp-wgsp tree single peaked}.}]
			We demonstrate the desired implications for the equivalence in turn.\medskip
			
			\noindent \textbf{\ref{item TD} $\implies$ \ref{item DA is wgsp}.} 
			This implication follows from Theorem \ref{theorem mandal result}.
			
			\noindent \textbf{\ref{item DA is wgsp} $\implies$ \ref{item stable and wgsp} $\implies$ \ref{item stable and sp}.} 
			These implications are straightforward.\medskip
			
			Notice that the above implications do not depend on richness, anonymity, or tree-single-peakedness.\medskip
			
			\noindent \textbf{\ref{item stable and sp} $\implies$ \ref{item TD}.} 
			We prove the contrapositive: suppose $\mathcal{P}_A$ satisfies top dominance for neither of the sides, then no stable matching rule on $\mathcal{P}_A$ is strategy-proof.
			
			Since $\mathcal{P}_A$ is a rich and anonymous tree-single-peaked subdomain that does not satisfy top dominance for either side of the market, by Lemma \ref{lemma TD on trees}, without loss of generality, there exist preferences $P_1, P_2, P_3 \in \mathcal{P}_{men}$ and $\tilde{P}_1, \tilde{P}_2, \tilde{P}_3 \in \mathcal{P}_{women}$ such that
			\begin{equation*}
				\begin{aligned}
					& w_2 \mathrel{P_1} w_1 \mathrel{P_1} w_3, \hspace{2 mm} w_2 \mathrel{P_2} w_3 \mathrel{P_2} w_1, \hspace{2 mm} w_1 \mathrel{P_3} w_2 \mathrel{P_3} w_3, \mbox{ and}\\
					& m_2 \mathrel{\tilde{P}_1} m_1 \mathrel{\tilde{P}_1} m_3, \hspace{2 mm} m_2 \mathrel{\tilde{P}_2} m_3 \mathrel{\tilde{P}_2} m_1, \hspace{2 mm} m_1 \mathrel{\tilde{P}_3} m_2 \mathrel{\tilde{P}_3} m_3.
				\end{aligned}
			\end{equation*}
			
			Since $\mathcal{P}_A$ is a rich and anonymous domain, we can construct the preference profiles presented in Table \ref{preference profiles one side has to be TD}. For instance, $w_k \cdots$ denotes a preference that ranks $w_k$ first (the dots indicate that all preferences for the corresponding parts are irrelevant and can be chosen arbitrarily). Here, $m_k$ denotes a man other than $m_1, m_2, m_3$ (if any), and $w_l$ denotes a woman other than $w_1, w_2, w_3$ (if any). Note that such an agent does not change their preference across the mentioned preference profiles.
			\begin{table}[H]
				\centering
				\begin{tabular}{@{}c|cccc|cccc@{}}
					\hline
					Preference profiles & $m_1$ & $m_2$ & $m_3$ & $m_k$ & $w_1$ & $w_2$ & $w_3$ & $w_l$ \\ \hline
					\hline
					$P^1_A$ & $P_3$ & $P_1$ & $P_3$ & $w_k \cdots$ & $\tilde{P}_1$ & $\tilde{P}_3$ & $\tilde{P}_2$ & $m_l \cdots$ \\
					$P^2_A$ & $P_3$ & $P_1$ & $P_3$ & $w_k \cdots$ & $\tilde{P}_2$ & $\tilde{P}_3$ & $\tilde{P}_2$ & $m_l \cdots$ \\
					$P^3_A$ & $P_3$ & $P_2$ & $P_3$ & $w_k \cdots$ & $\tilde{P}_1$ & $\tilde{P}_3$ & $\tilde{P}_2$ & $m_l \cdots$ \\
					\hline
				\end{tabular}
				\caption{Preference profiles for Theorem$^*$ \ref{theo sp-wgsp tree single peaked}}
				\label{preference profiles one side has to be TD}
			\end{table}
			
			It is straightforward to verify the following facts.
			\begin{enumerate}[(i)]
				\item \begin{enumerate}[(a)]
					\item $D^M(P^1_A) = [(m_1,w_1), (m_2,w_2), (m_k,w_k) \hspace{2 mm} \forall \hspace{2 mm} k \geq 3]$. We denote this matching by $\mu_1$ in this proof.
					
					\item $D^W(P^1_A) = [(m_1,w_2), (m_2,w_1), (m_k,w_k) \hspace{2 mm} \forall \hspace{2 mm} k \geq 3]$. We denote this matching by $\mu_2$ in this proof.
				\end{enumerate}
				
				\item $D^M(P^2_A) = D^W(P^2_A) = \mu_2$.
				
				\item $D^M(P^3_A) = D^W(P^3_A) = \mu_1$.
			\end{enumerate} 
			These facts, together with Remark \ref{remark properties of DA}, imply
			\begin{equation*}
				\mathcal{C}(P^1_A) = \{\mu_1, \mu_2\}, \hspace{2 mm} \mathcal{C}(P^2_A) = \{\mu_2\}, \mbox{ and } \mathcal{C}(P^3_A) = \{\mu_1\}, 
			\end{equation*}
			
			Fix a stable matching rule $\varphi$ on $\mathcal{P}_A$. Clearly, $\varphi(P^1_A) \in \{\mu_1, \mu_2\}$. If $\varphi(P^1_A) = \mu_1$, then $w_1$ can manipulate $\varphi$ at $P^1_A$ via $\tilde{P}_2$. If $\varphi(P^1_A) = \mu_2$, then $m_2$ can manipulate $\varphi$ at $P^1_A$ via $P_2$. Combining all these observations, it follows that $\varphi$ is not strategy-proof on $\mathcal{P}_A$, which completes the proof of Theorem$^*$ \ref{theo sp-wgsp tree single peaked}.
		\end{proof}

		\section{Proof of Proposition$^*$ \ref{proposition no stable and nb rule for atleast 5 tree}}\label{appendix proof of proposition no stable and nb rule for atleast 5 tree}
		
		\setcounter{equation}{0}
		\setcounter{table}{0}
		\setcounter{theorem}{0}
		\setcounter{definition}{0}
		\setcounter{obs}{0}
		\setcounter{remark}{0}
		\setcounter{example}{0}

		We first prove a lemma that we use in the proof of Proposition$^*$ \ref{proposition no stable and nb rule for atleast 5 tree}.

		\subsection{Lemma \ref{lemma existence only if rotation} and its proof}

		Lemma \ref{lemma existence only if rotation} identifies a necessary condition, called \textit{rotation}, of the domain for the existence of a stable and non-bossy matching rule. An anonymous domain satisfies \textit{rotation for women} if for every $m, \tilde{m}, m', \hat{m} \in M$ with $(\cdot m \cdot \tilde{m} \cdot m' \cdot \hat{m} \cdot) \in \mathcal{P}_{women}$, and every preference $P \in \mathcal{P}_{women}$ with $m' \mathrel{P} \hat{m}$, $m$ is either preferred to $m'$ or less preferred to $\hat{m}$ according to $P$. Below, we present the formal definition.

		\begin{definition}[Rotation]\label{definition rotation}
			An anonymous domain $\mathcal{P}_A$ satisfies \textit{\textbf{rotation} for women} if for every $m, \tilde{m}, m', \hat{m} \in M$,
			\begin{equation*}
				(\cdot m \cdot \tilde{m} \cdot m' \cdot \hat{m} \cdot) \in \mathcal{P}_{women} \implies (\cdot m' \cdot m \cdot \hat{m} \cdot) \notin \mathcal{P}_{women}.
			\end{equation*}
		\end{definition}

		We define \textit{\textbf{rotation} for men} in a similar way.
		
		We now state and prove Lemma \ref{lemma existence only if rotation}, which holds regardless of the domain being a (tree-)single-peaked subdomain.

		\begin{lemma}\label{lemma existence only if rotation}
			Suppose $\mathcal{P}_A$ is a rich and anonymous domain. Then, there exists a stable and non-bossy matching rule on $\mathcal{P}_A$ only if $\mathcal{P}_A$ satisfies rotation for at least one side of the market.
		\end{lemma}

		\begin{proof}[\textbf{Proof of Lemma \ref{lemma existence only if rotation}}]
			Suppose $\mathcal{P}_A$ satisfies rotation for neither of the sides. We show that there is no stable and non-bossy matching rule on $\mathcal{P}_A$.
			Since $\mathcal{P}_A$ does not satisfy rotation for either side of the market, without loss of generality, there exist preferences $P_1, P_2 \in \mathcal{P}_{men}$ and $\tilde{P}_1, \tilde{P}_2 \in \mathcal{P}_{women}$ such that
			\begin{equation*}
				\begin{aligned}
					& w_1 \mathrel{P_1} w_2 \mathrel{P_1} w_4 \hspace{1 mm} \mbox{ and } \hspace{1 mm} w_2 \mathrel{P_2} w_3 \mathrel{P_2} w_1 \mathrel{P_2} w_4, \hspace{1 mm} \mbox{ and}\\
					& m_1 \mathrel{\tilde{P}_1} m_2 \mathrel{\tilde{P}_1} m_4 \hspace{1 mm} \mbox{ and } \hspace{1 mm} m_2 \mathrel{\tilde{P}_2} m_3 \mathrel{\tilde{P}_2} m_1 \mathrel{\tilde{P}_2} m_4.
				\end{aligned}
			\end{equation*}
			We distinguish the following four cases.\medskip

			\noindent\textbf{\textsc{Case} 1}: Suppose $w_1 \mathrel{P_1} w_3$ and $m_1 \mathrel{\tilde{P}_1} m_3$.
			
			Since $\mathcal{P}_A$ is a rich domain, there exist four preferences $P_3, P_4 \in \mathcal{P}_{men}$ and $\tilde{P}_3, \tilde{P}_4 \in \mathcal{P}_{women}$ such that $\tau(P_3) = w_3$, $\tau(P_4) = w_4$, $\tau(\tilde{P}_3) = m_3$, and $\tau(\tilde{P}_4) = m_4$. Moreover, since $\mathcal{P}_A$ is a rich and anonymous domain, we can construct the preference profiles presented in Table \ref{preference profiles one side rotation case 1}. Here, $m_k$ denotes a man other than $m_1, m_2, m_3, m_4$ (if any), and $w_l$ denotes a woman other than $w_1, w_2, w_3, w_4$ (if any). Note that such an agent does not change their preference across the mentioned preference profiles.
			\begin{table}[H]
				\centering
				\begin{tabular}{@{}c|ccccc|ccccc@{}}
					\hline
					Preference profiles & $m_1$ & $m_2$ & $m_3$ & $m_4$ & $m_k$ & $w_1$ & $w_2$ & $w_3$ & $w_4$ & $w_l$ \\ \hline
					\hline
					$P^1_A$ & $P_1$ & $P_2$ & $P_4$ & $P_3$ & $w_k \cdots$ & $\tilde{P}_2$ & $\tilde{P}_1$ & $\tilde{P}_4$ & $\tilde{P}_3$ & $m_l \cdots$ \\
					$P^2_A$ & $P_1$ & $P_2$ & $P_1$ & $P_3$ & $w_k \cdots$ & $\tilde{P}_2$ & $\tilde{P}_1$ & $\tilde{P}_4$ & $\tilde{P}_3$ & $m_l \cdots$ \\
					$P^3_A$ & $P_1$ & $P_2$ & $P_4$ & $P_3$ & $w_k \cdots$ & $\tilde{P}_2$ & $\tilde{P}_1$ & $\tilde{P}_1$ & $\tilde{P}_3$ & $m_l \cdots$ \\
					\hline
				\end{tabular}
				\caption{Preference profiles for Case 1 of Lemma \ref{lemma existence only if rotation}}
				\label{preference profiles one side rotation case 1}
			\end{table}
			
			It is straightforward to verify the following facts.
			\begin{enumerate}[(i)]
				\item \begin{enumerate}[(a)]
					\item $D^M(P^1_A) = [(m_1,w_1), (m_2,w_2), (m_3,w_4), (m_4,w_3), (m_k,w_k) \hspace{2 mm} \forall \hspace{2 mm} k \geq 5]$. We denote this matching by $\mu_1$ in this proof.
					
					\item $D^W(P^1_A) = [(m_1,w_2), (m_2,w_1), (m_3,w_4), (m_4,w_3), (m_k,w_k) \hspace{2 mm} \forall \hspace{2 mm} k \geq 5]$. We denote this matching by $\mu_2$ in this proof.
				\end{enumerate}
				
				\item $D^M(P^2_A) = D^W(P^2_A) = \mu_2$.
				
				\item $D^M(P^3_A) = D^W(P^3_A) = \mu_1$.
			\end{enumerate} 
			These facts, together with Remark \ref{remark properties of DA}, imply
			\begin{equation*}
				\mathcal{C}(P^1_A) = \{\mu_1, \mu_2\}, \hspace{2 mm} \mathcal{C}(P^2_A) = \{\mu_2\}, \mbox{ and } \mathcal{C}(P^3_A) = \{\mu_1\}, 
			\end{equation*}
			Fix a stable matching rule $\varphi$ on $\mathcal{P}_A$. Clearly, $\varphi(P^1_A) \in \{\mu_1, \mu_2\}$.
			\begin{enumerate}[(i)]
				\item Suppose $\varphi(P^1_A) = \mu_1$. 
				
				Note that only $m_3$ changes his preference from $P^1_A$ to $P^2_A$. This, together with the facts $\varphi(P^1_A) = \mu_1$ and $\varphi(P^2_A) = \mu_2$, implies that $\varphi$ violates non-bossiness.
				
				\item Suppose $\varphi(P^1_A) = \mu_2$. 
				
				Note that only $w_3$ changes her preference from $P^1_A$ to $P^3_A$. This, together with the facts $\varphi(P^1_A) = \mu_2$ and $\varphi(P^3_A) = \mu_1$, implies that $\varphi$ violates non-bossiness.
			\end{enumerate}\medskip

			\noindent\textbf{\textsc{Case} 2}: Suppose $w_1 \mathrel{P_1} w_3$ and $m_3 \mathrel{\tilde{P}_1} m_1$.
			
			By renaming men $m_1, m_2, m_3$ as $m'_3, m'_1, m'_2$, respectively, and renaming preferences $\tilde{P}_1, \tilde{P}_2$ as $\tilde{P}'_2, \tilde{P}'_1$, respectively, we obtain an identical situation to Case 1.\medskip

			\noindent\textbf{\textsc{Case} 3}: Suppose $w_3 \mathrel{P_1} w_1$ and $m_1 \mathrel{\tilde{P}_1} m_3$.
			
			By renaming women $w_1, w_2, w_3$ as $w'_3, w'_1, w'_2$, respectively, and renaming preferences $P_1, P_2$ as $P'_2, P'_1$, respectively, we obtain an identical situation to Case 1.\medskip

			\noindent\textbf{\textsc{Case} 4}: Suppose $w_3 \mathrel{P_1} w_1$ and $m_3 \mathrel{\tilde{P}_1} m_1$.
			
			By renaming men $m_1, m_2, m_3$ as $m'_3, m'_1, m'_2$, respectively, renaming women $w_1, w_2, w_3$ as $w'_3, w'_1, w'_2$, respectively, and renaming preferences $P_1, P_2, \tilde{P}_1, \tilde{P}_2$ as $P'_2, P'_1, \tilde{P}'_2, \tilde{P}'_1$, respectively, we obtain an identical situation to Case 1.\medskip
			
			Since Cases 1 -- 4 are exhaustive, this completes the proof of Lemma \ref{lemma existence only if rotation}.
		\end{proof}

		\subsection{Completion of the proof of Proposition$^*$ \ref{proposition no stable and nb rule for atleast 5 tree}}

		Suppose $\mathcal{P}_A$ is a rich and anonymous tree-single-peaked subdomain, and let $n \geq 5$. We first prove that $\mathcal{P}_A$ does not satisfy rotation for either side of the market through two claims.

		\begin{claim}\label{claim no rotation for men}
			$\mathcal{P}_A$ does not satisfy rotation for men.
		\end{claim}

		\begin{claimproof}[Proof of Claim \ref{claim no rotation for men}]
			Consider a subtree $\tilde{T}_W$ of $T_W$ such that $|V(\tilde{T}_W)| = 5$. $\tilde{T}_W$ has exactly one of the three tree structures presented in Figure \ref{tree possible structures}. 
			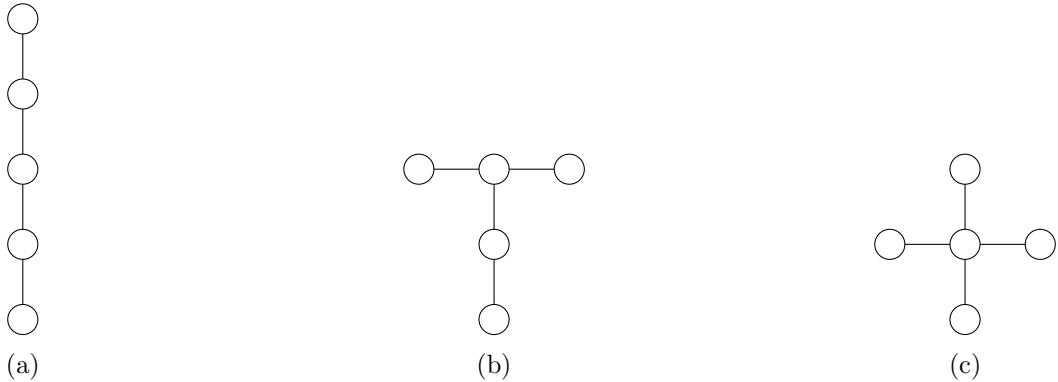
\begin{figure}[H]
				\centering
				\begin{subfigure}[b]{0.3\textwidth}
					\centering
					\begin{tikzpicture}[->]
						\begin{scope}
							\draw[draw = black] (1,1) circle (.2) (1,2) circle (.2) (1,3) circle (.2) (1,4) circle (.2) (1,5) circle (.2);
							\path (1,1.2) edge[-] node {} (1,1.8);
							\path (1,2.2) edge[-] node {} (1,2.8);
							\path (1,3.2) edge[-] node {} (1,3.8);
							\path (1,4.2) edge[-] node {} (1,4.8);
						\end{scope}
					\end{tikzpicture}
					\caption{}
					\label{figure tree type 1}
				\end{subfigure}
				\quad
				\begin{subfigure}[b]{0.3\textwidth}
					\centering
					\begin{tikzpicture}[->]
						\begin{scope}
							\draw[draw = black] (1,1) circle (.2) (1,2) circle (.2) (1,3) circle (.2) (0,3) circle (.2) (2,3) circle (.2);
							\path (1,1.2) edge[-] node {} (1,1.8);
							\path (1,2.2) edge[-] node {} (1,2.8);
							\path (0.2,3) edge[-] node {} (0.8,3);
							\path (1.2,3) edge[-] node {} (1.8,3);
						\end{scope}
					\end{tikzpicture}
					\caption{}
					\label{figure tree type 2}
				\end{subfigure}
				\quad
				\begin{subfigure}[b]{0.3\textwidth}
					\centering
					\begin{tikzpicture}[->]
						\begin{scope}
							\draw[draw = black] (1,1) circle (.2) (1,2) circle (.2) (0,2) circle (.2) (2,2) circle (.2) (1,3) circle (.2);
							\path (1,1.2) edge[-] node {} (1,1.8);
							\path (1,2.2) edge[-] node {} (1,2.8);
							\path (0.2,2) edge[-] node {} (0.8,2);
							\path (1.2,2) edge[-] node {} (1.8,2);
						\end{scope}
					\end{tikzpicture}
					\caption{}
					\label{figure tree type 3}
				\end{subfigure}
				\caption{Possible tree structures for $\tilde{T}_W$}
				\label{tree possible structures}
			\end{figure}

			We distinguish the following three cases.\medskip

			\noindent\textbf{\textsc{Case} 1}: Suppose $\tilde{T}_W$ has the tree structure given in Figure \ref{figure tree type 1}. 
			
			Without loss of generality, assume that $V(\tilde{T}_W) = \{w_1, w_2, w_3, w_4, w_5\}$ and $\tilde{T}_W$ is as given in Figure \ref{figure tree structure for case 1}.
			\begin{figure}[H] 
				\centering
				\begin{tikzpicture}[->]
					\begin{scope}
						\draw[draw = black] (1,1) circle (.3) (3,1) circle (.3) (5,1) circle (.3) (7,1) circle (.3) (9,1) circle (.3);
						\node at (1,1) {$w_1$}; \node at (3,1) {$w_2$}; \node at (5,1) {$w_3$}; \node at (7,1) {$w_4$}; \node at (9,1) {$w_5$};
						\path (1.3,1) edge[-] node {} (2.7,1);
						\path (3.3,1) edge[-] node {} (4.7,1);
						\path (5.3,1) edge[-] node {} (6.7,1);
						\path (7.3,1) edge[-] node {} (8.7,1);
					\end{scope}
				\end{tikzpicture}
				\caption{Subtree $\tilde{T}_W$ for Case 1}
				\label{figure tree structure for case 1}
			\end{figure}

			Because $\mathcal{P}_A$ is rich and anonymous, there exist preferences $P_1, P_3, P_5 \in \mathcal{P}_{men}$ such that $\tau(P_1) = w_1$, $\tau(P_3) = w_3$, and $\tau(P_5) = w_5$. Furthermore, since $\mathcal{P}_{men} \subseteq \mathbb{S}(T_W)$ and $\tilde{T}_W$ is a subtree of $T_W$, the facts $\tau(P_1) = w_1$ and $\tau(P_5) = w_5$, together with the structure of $\tilde{T}_W$, imply
			\begin{subequations}\label{equation impossibility 1}
				\begin{equation}\label{equation impossibility 1 a}
					w_1 \mathrel{P_1} w_2 \mathrel{P_1} w_3 \mathrel{P_1} w_5, \mbox{ and}
				\end{equation}
				\begin{equation}\label{equation impossibility 1 b}
					w_5 \mathrel{P_5} w_4 \mathrel{P_5} w_3 \mathrel{P_5} w_1.
				\end{equation}
			\end{subequations}
			
			\begin{enumerate}[(i)]
				\item Suppose $w_3 \mathrel{P_3} w_1 \mathrel{P_3} w_5$.
				
				The fact $w_3 \mathrel{P_3} w_1 \mathrel{P_3} w_5$ and \eqref{equation impossibility 1 a} together imply that $\mathcal{P}_A$ does not satisfy rotation for men.
				
				\item Suppose $w_3 \mathrel{P_3} w_5 \mathrel{P_3} w_1$.
				
				The fact $w_3 \mathrel{P_3} w_5 \mathrel{P_3} w_1$ and \eqref{equation impossibility 1 b} together imply that $\mathcal{P}_A$ does not satisfy rotation for men.
			\end{enumerate}\medskip

			\noindent\textbf{\textsc{Case} 2}: Suppose $\tilde{T}_W$ has the tree structure given in Figure \ref{figure tree type 2}. 
			
			Without loss of generality, assume that $V(\tilde{T}_W) = \{w_1, w_2, w_3, w_4, w_5\}$ and $\tilde{T}_W$ is as given in Figure \ref{figure tree structure for case 2}.
			\begin{figure}[H] 
				\centering
				\begin{tikzpicture}[->]
					\begin{scope}
						\draw[draw = black] (1,1) circle (.3) (3,1) circle (.3) (5,1) circle (.3) (7,2) circle (.3) (7,0) circle (.3);
						\node at (1,1) {$w_1$}; \node at (3,1) {$w_2$}; \node at (5,1) {$w_3$}; \node at (7,2) {$w_4$}; \node at (7,0) {$w_5$};
						\path (1.3,1) edge[-] node {} (2.7,1);
						\path (3.3,1) edge[-] node {} (4.7,1);
						\path (5.25,1.15) edge[-] node {} (6.75,1.85);
						\path (5.25,.85) edge[-] node {} (6.75,.15);
					\end{scope}
				\end{tikzpicture}
				\caption{Subtree $\tilde{T}_W$ for Case 2}
				\label{figure tree structure for case 2}
			\end{figure}

			Because $\mathcal{P}_A$ is rich and anonymous, there exist preferences $P_1, P_4, P_5 \in \mathcal{P}_{men}$ such that $\tau(P_1) = w_1$, $\tau(P_4) = w_4$, and $\tau(P_5) = w_5$. Furthermore, since $\mathcal{P}_{men} \subseteq \mathbb{S}(T_W)$ and $\tilde{T}_W$ is a subtree of $T_W$, the fact $\tau(P_1) = w_1$, together with the structure of $\tilde{T}_W$, implies
			\begin{equation}\label{equation impossibility 2}
				w_1 \mathrel{P_1} w_2 \mathrel{P_1} w_3 \mathrel{P_1} w_4 \mathrel{P_1} w_5 \hspace{2 mm} \mbox{ or } \hspace{2 mm} w_1 \mathrel{P_1} w_2 \mathrel{P_1} w_3 \mathrel{P_1} w_5 \mathrel{P_1} w_4.
			\end{equation}
			
			\begin{enumerate}[(i)]
				\item Suppose $w_1 \mathrel{P_1} w_2 \mathrel{P_1} w_3 \mathrel{P_1} w_4 \mathrel{P_1} w_5$.
				
				Assume for contradiction that $\mathcal{P}_A$ satisfies rotation for men. Since $\mathcal{P}_A$ is anonymous and satisfies rotation for men, the fact $w_2 \mathrel{P_1} w_3 \mathrel{P_1} w_4 \mathrel{P_1} w_5$ implies $(\cdot w_4 \cdot w_2 \cdot w_5 \cdot) \notin \mathcal{P}_{men}$. This, together with the fact $\tau(P_4) = w_4$ and the structure of $\tilde{T}_W$, implies $w_4 \mathrel{P_4} w_3 \mathrel{P_4} w_5 \mathrel{P_4} w_2 \mathrel{P_4} w_1$. Moreover, since $\mathcal{P}_A$ is anonymous and satisfies rotation for men, the fact $w_4 \mathrel{P_4} w_3 \mathrel{P_4} w_5 \mathrel{P_4} w_1$ implies $(\cdot w_5 \cdot w_4 \cdot w_1 \cdot) \notin \mathcal{P}_{men}$. This, together with the fact $\tau(P_5) = w_5$ and the structure of $\tilde{T}_W$, implies $w_5 \mathrel{P_5} w_3 \mathrel{P_5} w_2 \mathrel{P_5} w_1 \mathrel{P_5} w_4$. However, $w_1 \mathrel{P_1} w_3 \mathrel{P_1} w_4$ and $w_3 \mathrel{P_5} w_2 \mathrel{P_5} w_1 \mathrel{P_5} w_4$ together imply that $\mathcal{P}_A$ does not satisfy rotation for men.
				
				\item Suppose $w_1 \mathrel{P_1} w_2 \mathrel{P_1} w_3 \mathrel{P_1} w_5 \mathrel{P_1} w_4$.
				
				Assume for contradiction that $\mathcal{P}_A$ satisfies rotation for men. Since $\mathcal{P}_A$ is anonymous and satisfies rotation for men, the fact $w_2 \mathrel{P_1} w_3 \mathrel{P_1} w_5 \mathrel{P_1} w_4$ implies $(\cdot w_5 \cdot w_2 \cdot w_4 \cdot) \notin \mathcal{P}_{men}$. This, together with the fact $\tau(P_5) = w_5$ and the structure of $\tilde{T}_W$, implies $w_5 \mathrel{P_5} w_3 \mathrel{P_5} w_4 \mathrel{P_5} w_2 \mathrel{P_5} w_1$. Moreover, since $\mathcal{P}_A$ is anonymous and satisfies rotation for men, the fact $w_5 \mathrel{P_5} w_3 \mathrel{P_5} w_4 \mathrel{P_5} w_1$ implies $(\cdot w_4 \cdot w_5 \cdot w_1 \cdot) \notin \mathcal{P}_{men}$. This, together with the fact $\tau(P_4) = w_4$ and the structure of $\tilde{T}_W$, implies $w_4 \mathrel{P_4} w_3 \mathrel{P_4} w_2 \mathrel{P_4} w_1 \mathrel{P_4} w_5$. However, $w_1 \mathrel{P_1} w_3 \mathrel{P_1} w_5$ and $w_3 \mathrel{P_4} w_2 \mathrel{P_4} w_1 \mathrel{P_4} w_5$ together imply that $\mathcal{P}_A$ does not satisfy rotation for men.
			\end{enumerate}\medskip

			\noindent\textbf{\textsc{Case} 3}: Suppose $\tilde{T}_W$ has the tree structure given in Figure \ref{figure tree type 3}. 
			
			Without loss of generality, assume that $V(\tilde{T}_W) = \{w_1, w_2, w_3, w_4, w_5\}$ and $\tilde{T}_W$ is as given in Figure \ref{figure tree structure for case 3}.
			\begin{figure}[H] 
				\centering
				\begin{tikzpicture}[->]
					\begin{scope}
						\draw[draw = black] (1,1) circle (.3) (1,2) circle (.3) (0,2) circle (.3) (2,2) circle (.3) (1,3) circle (.3);
						\node at (1,1) {$w_2$}; \node at (1,2) {$w_1$}; \node at (0,2) {$w_3$}; \node at (2,2) {$w_5$}; \node at (1,3) {$w_4$};
						\path (1,1.3) edge[-] node {} (1,1.7);
						\path (1,2.3) edge[-] node {} (1,2.7);
						\path (0.3,2) edge[-] node {} (0.7,2);
						\path (1.3,2) edge[-] node {} (1.7,2);
					\end{scope}
				\end{tikzpicture}
				\caption{Subtree $\tilde{T}_W$ for Case 3}
				\label{figure tree structure for case 3}
			\end{figure}
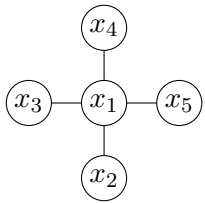

			Because $\mathcal{P}_A$ is rich and anonymous, there exist a preference $P_1 \in \mathcal{P}_{men}$ such that $\tau(P_1) = w_1$. Notice that $w_2$, $w_3$, $w_4$, and $w_5$ can appear in any order in $P_1$ after $w_1$. Without loss of generality, assume that $w_1 \mathrel{P_1} w_2 \mathrel{P_1} w_3 \mathrel{P_1} w_4 \mathrel{P_1} w_5$.
			
			Moreover, because $\mathcal{P}_A$ is rich and anonymous, there exist a preference $P_3 \in \mathcal{P}_{men}$ such that $\tau(P_3) = w_3$. Since $\mathcal{P}_{men} \subseteq \mathbb{S}(T_W)$ and $\tilde{T}_W$ is a subtree of $T_W$, the fact $\tau(P_3) = w_3$, together with the structure of $\tilde{T}_W$, implies $w_3 \mathrel{P_3} w_1 \mathrel{P_3} w_4$. However, this, along with the fact $w_1 \mathrel{P_1} w_2 \mathrel{P_1} w_3 \mathrel{P_1} w_4$, implies that $\mathcal{P}_A$ does not satisfy rotation for men.\medskip
			
			Since Cases 1 -- 3 are exhaustive, this completes the proof of Claim \ref{claim no rotation for men}
		\end{claimproof}

		\begin{claim}\label{claim no rotation for women}
			$\mathcal{P}_A$ does not satisfy rotation for women.
		\end{claim}
		
		\begin{claimproof}[Proof of Claim \ref{claim no rotation for women}]
			Proof of Claim \ref{claim no rotation for women} follows using a similar argument as for Claim \ref{claim no rotation for men}.
		\end{claimproof}

		We now proceed to complete the proof of Proposition$^*$ \ref{proposition no stable and nb rule for atleast 5 tree}. By Claims \ref{claim no rotation for men} and \ref{claim no rotation for women}, $\mathcal{P}_A$ does not satisfy rotation for either side of the market, and consequently, Proposition$^*$ \ref{proposition no stable and nb rule for atleast 5 tree} follows from Lemma \ref{lemma existence only if rotation}.
		\hfill
		\qed

		\section{A stable matching rule which is strategy-proof but not weakly group strategy-proof}\label{appendix example}
		
		\setcounter{equation}{0}
		\setcounter{table}{0}
		\setcounter{theorem}{0}
		\setcounter{definition}{0}
		\setcounter{obs}{0}
		\setcounter{remark}{0}
		\setcounter{example}{0}

		As discussed in Section \ref{section results}, Theorem \ref{theo sp-wgsp classical single peaked} demonstrates that under richness and anonymity, the (classical) single-peaked subdomains on which stability and strategy-proofness are compatible are the same ones on which stability and weak group strategy-proofness are compatible. However, not every stable and strategy-proof matching rule on such a domain is weakly group strategy-proof. An example is provided below to illustrate this.

		\begin{example}\label{example SP but not WGSP}
			Suppose $M = \{m_1, \ldots, m_6\}$ and $W = \{w_1, \ldots, w_6\}$ with prior (linear) orderings $\prec_M$ and $\prec_W$, respectively, such that $m_1 \prec_M \ldots \prec_M m_6$ and $w_1 \prec_W \ldots \prec_W w_6$. 
			Consider the rich and anonymous single-peaked subdomain $\mathcal{P}_A$ as follows:
			\begin{equation*}
				\begin{aligned}
					& \mathcal{P}_{men} = \left\{ 
					\begin{aligned}
						& w_1 w_2 w_3 w_4 w_5 w_6, w_2 w_3 w_4 w_5 w_1 w_6, w_2 w_3 w_4 w_5 w_6 w_1, w_3 w_2 w_4 w_5 w_1 w_6,\\
						& w_4 w_5 w_3 w_2 w_6 w_1, w_5 w_4 w_3 w_2 w_1 w_6, w_5 w_4 w_3 w_2 w_6 w_1, w_6 w_5 w_4 w_3 w_2 w_1 
					\end{aligned}
					\right\} \mbox{ and}\\
					& \mathcal{P}_{women} = \left\{ 
					\begin{aligned}
						& m_1 m_2 m_3 m_4 m_5 m_6, m_2 m_3 m_4 m_5 m_6 m_1, m_3 m_2 m_4 m_5 m_6 m_1,\\
						& m_4 m_5 m_3 m_2 m_6 m_1, m_5 m_4 m_3 m_2 m_6 m_1, m_6 m_5 m_4 m_3 m_2 m_1
					\end{aligned}
					\right\}.
				\end{aligned}
			\end{equation*} 
			By construction, $\mathcal{P}_A$ satisfies top dominance for women, and consequently, by Theorem \ref{theorem mandal result}, the matching rule $D^M$ is stable and weakly group strategy-proof on $\mathcal{P}_A$.
			
			We now construct a stable matching rule $\varphi$ on $\mathcal{P}_A$, and show that $\varphi$ is strategy-proof on $\mathcal{P}_A$ but not weakly group strategy-proof. For ease of presentation, we use the following notations for the construction:
			\begin{enumerate}[nolistsep]
				\item $\mu_1$ denotes the matching $[(m_1,w_1), (m_2,w_2), (m_3,w_3), (m_4,w_4), (m_5,w_5), (m_6,w_6)]$.
				
				\item $\mu_2$ denotes the matching $[(m_1,w_1), (m_2,w_2), (m_3,w_3), (m_4,w_5), (m_5,w_4), (m_6,w_6)]$.
				
				\item $\mu_3$ denotes the matching $[(m_1,w_1), (m_2,w_3), (m_3,w_2), (m_4,w_4), (m_5,w_5), (m_6,w_6)]$.
				
				\item $\mu_4$ denotes the matching $[(m_1,w_1), (m_2,w_3), (m_3,w_2), (m_4,w_5), (m_5,w_4), (m_6,w_6)]$.
			\end{enumerate}
			Furthermore, consider the following two subdomains $\mathcal{\tilde{P}}_A$ and $\mathcal{\hat{P}}_A$ of $\mathcal{P}_A$ such that
			\begin{equation*}
				\mathcal{\tilde{P}}_A = \left\{ \tilde{P}_A : 
				\begin{aligned}
					& \tilde{P}_{m_1} = w_1 w_2 w_3 w_4 w_5 w_6, \tilde{P}_{m_2} = w_2 w_3 w_4 w_5 w_1 w_6, \tilde{P}_{m_3} = w_3 w_2 w_4 w_5 w_1 w_6,\\
					& \tilde{P}_{m_4} \in \{w_1 w_2 w_3 w_4 w_5 w_6, w_2 w_3 w_4 w_5 w_1 w_6, w_2 w_3 w_4 w_5 w_6 w_1, w_3 w_2 w_4 w_5 w_1 w_6, w_4 w_5 w_3 w_2 w_6 w_1\},\\
					& \tilde{P}_{m_5} \in \{w_5 w_4 w_3 w_2 w_1 w_6, w_5 w_4 w_3 w_2 w_6 w_1, w_6 w_5 w_4 w_3 w_2 w_1\},\\
					& \tilde{P}_{m_6} = w_6 w_5 w_4 w_3 w_2 w_1,\\
					& \tilde{P}_{w_1} = m_1 m_2 m_3 m_4 m_5 m_6, \tilde{P}_{w_2} = m_3 m_2 m_4 m_5 m_6 m_1, \tilde{P}_{w_3} = m_2 m_3 m_4 m_5 m_6 m_1,\\
					& \tilde{P}_{w_4} \in \{m_5 m_4 m_3 m_2 m_6 m_1, m_6 m_5 m_4 m_3 m_2 m_1\},\\
					& \tilde{P}_{w_5} \in \{m_1 m_2 m_3 m_4 m_5 m_6, m_2 m_3 m_4 m_5 m_6 m_1, m_3 m_2 m_4 m_5 m_6 m_1, m_4 m_5 m_3 m_2 m_6 m_1\},\\
					& \tilde{P}_{w_6} = m_6 m_5 m_4 m_3 m_2 m_1
				\end{aligned}
				\right\}
			\end{equation*}
			and
			\begin{equation*}
				\mathcal{\hat{P}}_A = \left\{ \hat{P}_A :
				\begin{aligned}
					& \hat{P}_{m_1} = w_1 w_2 w_3 w_4 w_5 w_6,\\
					& \hat{P}_{m_2} \in \{w_1 w_2 w_3 w_4 w_5 w_6, w_2 w_3 w_4 w_5 w_1 w_6, w_2 w_3 w_4 w_5 w_6 w_1\},\\
					& \hat{P}_{m_3} \in \{w_3 w_2 w_4 w_5 w_1 w_6, w_4 w_5 w_3 w_2 w_6 w_1, w_5 w_4 w_3 w_2 w_1 w_6, w_5 w_4 w_3 w_2 w_6 w_1, w_6 w_5 w_4 w_3 w_2 w_1\},\\
					& \hat{P}_{m_4} = w_4 w_5 w_3 w_2 w_6 w_1, \hat{P}_{m_5} = w_5 w_4 w_3 w_2 w_6 w_1, \hat{P}_{m_6} = w_6 w_5 w_4 w_3 w_2 w_1,\\
					& \hat{P}_{w_1} = m_1 m_2 m_3 m_4 m_5 m_6,\\
					& \hat{P}_{w_2} \in \{m_3 m_2 m_4 m_5 m_6 m_1, m_4 m_5 m_3 m_2 m_6 m_1, m_5 m_4 m_3 m_2 m_6 m_1, m_6 m_5 m_4 m_3 m_2 m_1\},\\
					& \hat{P}_{w_3} \in \{m_1 m_2 m_3 m_4 m_5 m_6, m_2 m_3 m_4 m_5 m_6 m_1\},\\
					& \hat{P}_{w_4} = m_5 m_4 m_3 m_2 m_6 m_1, \hat{P}_{w_5} = m_4 m_5 m_3 m_2 m_6 m_1, \hat{P}_{w_6} = m_6 m_5 m_4 m_3 m_2 m_1
				\end{aligned}
				\right\}.
			\end{equation*}
			By construction, $\mathcal{\tilde{P}}_A$ and $\mathcal{\hat{P}}_A$ have exactly one common preference profile -- let us denote it by $P^*_A$ -- where
			\begin{equation*}
				\begin{aligned}
					& P^*_{m_1} = w_1 w_2 w_3 w_4 w_5 w_6, P^*_{m_2} = w_2 w_3 w_4 w_5 w_1 w_6, P^*_{m_3} = w_3 w_2 w_4 w_5 w_1 w_6,\\
					& P^*_{m_4} = w_4 w_5 w_3 w_2 w_6 w_1, P^*_{m_5} = w_5 w_4 w_3 w_2 w_6 w_1, P^*_{m_6} = w_6 w_5 w_4 w_3 w_2 w_1,\\
					& P^*_{w_1} = m_1 m_2 m_3 m_4 m_5 m_6, P^*_{w_2} = m_3 m_2 m_4 m_5 m_6 m_1, P^*_{w_3} = m_2 m_3 m_4 m_5 m_6 m_1,\\
					& P^*_{w_4} = m_5 m_4 m_3 m_2 m_6 m_1, P^*_{w_5} = m_4 m_5 m_3 m_2 m_6 m_1, P^*_{w_6} = m_6 m_5 m_4 m_3 m_2 m_1.
				\end{aligned}
			\end{equation*} 
			
			We now construct the matching rule $\varphi$ on $\mathcal{P}_A$ as follows:\\
			
			$\varphi(P_A) = \begin{cases}
				\mu_2 & \mbox{ if } P_A \in \mathcal{\tilde{P}}_A \setminus \{P^*_A\}\\
				\mu_3 & \mbox{ if } P_A \in \mathcal{\hat{P}}_A \setminus \{P^*_A\}\\
				\mu_4 & \mbox{ if } P_A = P^*_A\\
				D^M(P_A) & \mbox{ if } P_A \in \mathcal{P}_A \setminus (\mathcal{\tilde{P}}_A \cup \mathcal{\hat{P}}_A)
			\end{cases}$\\\\
			By construction, $\varphi$ is stable on $\mathcal{P}_A$. To see this, note that $\mathcal{C}(P_A) = \{\mu_1, \mu_2, \mu_3, \mu_4\}$ for all $P_A \in \mathcal{\tilde{P}}_A \cup \mathcal{\hat{P}}_A$.
			
			We first show that $\varphi$ does not satisfy weak group strategy-proofness on $\mathcal{P}_A$. Let $A'$ be a coalition of men $m_2$ and $m_5$. Consider the preference profile $P'_{A'}$ of this coalition such that
			\begin{equation*}
				P'_{m_2} = w_2 w_3 w_4 w_5 w_6 w_1 \hspace{2 mm} \mbox{ and } \hspace{2 mm} P'_{m_5} = w_5 w_4 w_3 w_2 w_1 w_6.
			\end{equation*}
			By construction, $(P'_{A'}, P^*_{-A'}) \in \mathcal{P}_A \setminus (\mathcal{\tilde{P}}_A \cup \mathcal{\hat{P}}_A)$, and therefore, $\varphi(P'_{A'}, P^*_{-A'}) = \mu_1$. This, along with the fact $\varphi(P^*_A) = \mu_4$, implies that the coalition $A'$ manipulates $\varphi$ at $P^*_A$ via $P'_{A'}$.
			
			We now show that $\varphi$ is strategy-proof on $\mathcal{P}_A$. Assume for contradiction that this is not true; there exist a preference profile $\bar{P}_A \in \mathcal{P}_A$, an agent $a \in A$, and a preference $\bar{P}'_a \in \mathcal{P}_a$ of agent $a$ such that agent $a$ manipulates $\varphi$ at $\bar{P}_A$ via $\bar{P}'_a$. We distinguish the following nine cases.
			\begin{enumerate}[(i)]
				\item Suppose $a \in \{m_1, m_6, w_1, w_6\}$.
				
				Since $a \in \{m_1, m_6, w_1, w_6\}$, it follows from the construction of $\varphi$ that $\varphi_a(P_A) = D^M_a(P_A)$ for all $P_A \in \mathcal{P}_A$. Moreover, because agent $a$ manipulates $\varphi$ at $\bar{P}_A$ via $\bar{P}'_a$, this implies that agent $a$ manipulates $D^M$ at $\bar{P}_A$ via $\bar{P}'_a$, which contradicts the fact that $D^M$ satisfies (weak group) strategy-proofness on $\mathcal{P}_A$.

				\item\label{case m2 manipulator} Suppose $a = m_2$.
				
				It follows from the construction of $\varphi$ that
				\begin{subequations}\label{equation m2 manipulative}
					\begin{equation}\label{equation m2 manipulative 1}
						\varphi_{m_2}(P_A) = w_3 \mbox{ for all } P_A \in \mathcal{\hat{P}}_A, \mbox{ and}
					\end{equation}
					\begin{equation}\label{equation m2 manipulative 2}
						\varphi_{m_2}(P_A) = D^M_{m_2}(P_A) \mbox{ for all } P_A \in \mathcal{P}_A \setminus \mathcal{\hat{P}}_A.
					\end{equation}
				\end{subequations}
				Since $m_2$ manipulates $\varphi$ at $\bar{P}_A$ via $\bar{P}'_{m_2}$, \eqref{equation m2 manipulative 1} implies that both $\bar{P}_A$ and $(\bar{P}'_{m_2}, \bar{P}_{-m_2})$ cannot be in $\mathcal{\hat{P}}_A$. Moreover, since $m_2$ manipulates $\varphi$ at $\bar{P}_A$ via $\bar{P}'_{m_2}$ and because $D^M$ satisfies (weak group) strategy-proofness on $\mathcal{P}_A$, \eqref{equation m2 manipulative 2} implies that both $\bar{P}_A$ and $(\bar{P}'_{m_2}, \bar{P}_{-m_2})$ cannot be in $\mathcal{P}_A \setminus \mathcal{\hat{P}}_A$ either. Combining these two observations, it follows that one of $\bar{P}_A$ and $(\bar{P}'_{m_2}, \bar{P}_{-m_2})$ is in $\mathcal{\hat{P}}_A$, and the other is in $\mathcal{P}_A \setminus \mathcal{\hat{P}}_A$.
				
				Suppose $\bar{P}_A \in \mathcal{\hat{P}}_A$ and $(\bar{P}'_{m_2}, \bar{P}_{-m_2}) \in \mathcal{P}_A \setminus \mathcal{\hat{P}}_A$. Then, $\varphi_{m_2}(\bar{P}_A) = w_3$, and because of $\bar{P}_{-m_2}$'s structure, we have $\varphi_{m_2}(\bar{P}'_{m_2}, \bar{P}_{-m_2}) = w_3$. However, the facts $\varphi_{m_2}(\bar{P}_A) = w_3$ and $\varphi_{m_2}(\bar{P}'_{m_2}, \bar{P}_{-m_2}) = w_3$ together contradict that $m_2$ manipulates $\varphi$ at $\bar{P}_A$ via $\bar{P}'_{m_2}$.
				
				Suppose $\bar{P}_A \in \mathcal{P}_A \setminus \mathcal{\hat{P}}_A$ and $(\bar{P}'_{m_2}, \bar{P}_{-m_2}) \in \mathcal{\hat{P}}_A$. Then, $\varphi_{m_2}(\bar{P}'_{m_2}, \bar{P}_{-m_2}) = w_3$, and because of $\bar{P}_{-m_2}$'s structure, we have $\varphi_{m_2}(\bar{P}_A) = w_3$. However, the facts $\varphi_{m_2}(\bar{P}_A) = w_3$ and $\varphi_{m_2}(\bar{P}'_{m_2}, \bar{P}_{-m_2}) = w_3$ together contradict that $m_2$ manipulates $\varphi$ at $\bar{P}_A$ via $\bar{P}'_{m_2}$.

				\item Suppose $a = m_3$.
				
				It follows from the construction of $\varphi$ that
				\begin{subequations}\label{equation m3 manipulative}
					\begin{equation}\label{equation m3 manipulative 1}
						\varphi_{m_3}(P_A) = w_2 \mbox{ for all } P_A \in \mathcal{\hat{P}}_A, \mbox{ and}
					\end{equation}
					\begin{equation}\label{equation m3 manipulative 2}
						\varphi_{m_3}(P_A) = D^M_{m_3}(P_A) \mbox{ for all } P_A \in \mathcal{P}_A \setminus \mathcal{\hat{P}}_A.
					\end{equation}
				\end{subequations}
				Using a similar argument as for Case \ref{case m2 manipulator} (and using \eqref{equation m3 manipulative} instead of \eqref{equation m2 manipulative}), it follows that one of $\bar{P}_A$ and $(\bar{P}'_{m_3}, \bar{P}_{-m_3})$ is in $\mathcal{\hat{P}}_A$, and the other is in $\mathcal{P}_A \setminus \mathcal{\hat{P}}_A$. Moreover, it also follows that $\varphi_{m_3}(\bar{P}_A) = \varphi_{m_3}(\bar{P}'_{m_3}, \bar{P}_{-m_3}) = w_2$, which contradicts the fact that $m_3$ manipulates $\varphi$ at $\bar{P}_A$ via $\bar{P}'_{m_3}$.

				\item Suppose $a = m_4$.
				
				It follows from the construction of $\varphi$ that
				\begin{subequations}\label{equation m4 manipulative}
					\begin{equation}\label{equation m4 manipulative 1}
						\varphi_{m_4}(P_A) = w_5 \mbox{ for all } P_A \in \mathcal{\tilde{P}}_A, \mbox{ and}
					\end{equation}
					\begin{equation}\label{equation m4 manipulative 2}
						\varphi_{m_4}(P_A) = D^M_{m_4}(P_A) \mbox{ for all } P_A \in \mathcal{P}_A \setminus \mathcal{\tilde{P}}_A.
					\end{equation}
				\end{subequations}
				Using a similar argument as for Case \ref{case m2 manipulator} (and using \eqref{equation m4 manipulative} instead of \eqref{equation m2 manipulative}), it follows that one of $\bar{P}_A$ and $(\bar{P}'_{m_4}, \bar{P}_{-m_4})$ is in $\mathcal{\tilde{P}}_A$, and the other is in $\mathcal{P}_A \setminus \mathcal{\tilde{P}}_A$. Moreover, it also follows that $\varphi_{m_4}(\bar{P}_A) = \varphi_{m_4}(\bar{P}'_{m_4}, \bar{P}_{-m_4}) = w_5$, which contradicts the fact that $m_4$ manipulates $\varphi$ at $\bar{P}_A$ via $\bar{P}'_{m_4}$.

				\item Suppose $a = m_5$.
				
				It follows from the construction of $\varphi$ that
				\begin{subequations}\label{equation m5 manipulative}
					\begin{equation}\label{equation m5 manipulative 1}
						\varphi_{m_5}(P_A) = w_4 \mbox{ for all } P_A \in \mathcal{\tilde{P}}_A, \mbox{ and}
					\end{equation}
					\begin{equation}\label{equation m5 manipulative 2}
						\varphi_{m_5}(P_A) = D^M_{m_5}(P_A) \mbox{ for all } P_A \in \mathcal{P}_A \setminus \mathcal{\tilde{P}}_A.
					\end{equation}
				\end{subequations}
				Using a similar argument as for Case \ref{case m2 manipulator} (and using \eqref{equation m5 manipulative} instead of \eqref{equation m2 manipulative}), it follows that one of $\bar{P}_A$ and $(\bar{P}'_{m_5}, \bar{P}_{-m_5})$ is in $\mathcal{\tilde{P}}_A$, and the other is in $\mathcal{P}_A \setminus \mathcal{\tilde{P}}_A$. Moreover, it also follows that $\varphi_{m_5}(\bar{P}_A) = \varphi_{m_5}(\bar{P}'_{m_5}, \bar{P}_{-m_5}) = w_4$, which contradicts the fact that $m_5$ manipulates $\varphi$ at $\bar{P}_A$ via $\bar{P}'_{m_5}$.

				\item\label{case w2 manipulator} Suppose $a = w_2$.
				
				It follows from the construction of $\varphi$ that
				\begin{subequations}\label{equation w2 manipulative}
					\begin{equation}\label{equation w2 manipulative 1}
						\varphi_{w_2}(P_A) = m_3 \mbox{ for all } P_A \in \mathcal{\hat{P}}_A, \mbox{ and}
					\end{equation}
					\begin{equation}\label{equation w2 manipulative 2}
						\varphi_{w_2}(P_A) = D^M_{w_2}(P_A) \mbox{ for all } P_A \in \mathcal{P}_A \setminus \mathcal{\hat{P}}_A.
					\end{equation}
				\end{subequations}
				Using a similar argument as for Case \ref{case m2 manipulator} (and using \eqref{equation w2 manipulative} instead of \eqref{equation m2 manipulative}), it follows that one of $\bar{P}_A$ and $(\bar{P}'_{w_2}, \bar{P}_{-w_2})$ is in $\mathcal{\hat{P}}_A$, and the other is in $\mathcal{P}_A \setminus \mathcal{\hat{P}}_A$. 
				
				Suppose $\bar{P}_A \in \mathcal{\hat{P}}_A$ and $(\bar{P}'_{w_2}, \bar{P}_{-w_2}) \in \mathcal{P}_A \setminus \mathcal{\hat{P}}_A$. Then, $\varphi_{w_2}(\bar{P}_A) = m_3$, and because of $\bar{P}_{-w_2}$'s structure, we have $\varphi_{w_2}(\bar{P}'_{w_2}, \bar{P}_{-w_2}) = m_2$. However, the facts $\varphi_{w_2}(\bar{P}_A) = m_3$ and $\varphi_{w_2}(\bar{P}'_{w_2}, \bar{P}_{-w_2}) = m_2$, along with $\bar{P}_{w_2}$'s structure, together imply $\varphi_{w_2}(\bar{P}_A) \mathrel{\bar{P}_{w_2}} \varphi_{w_2}(\bar{P}'_{w_2}, \bar{P}_{-w_2})$, which contradicts the fact that $w_2$ manipulates $\varphi$ at $\bar{P}_A$ via $\bar{P}'_{w_2}$.
				
				Suppose $\bar{P}_A \in \mathcal{P}_A \setminus \mathcal{\hat{P}}_A$ and $(\bar{P}'_{w_2}, \bar{P}_{-w_2}) \in \mathcal{\hat{P}}_A$. Then, $\varphi_{w_2}(\bar{P}'_{w_2}, \bar{P}_{-w_2}) = m_3$, and because of $\bar{P}_{-w_2}$'s structure, we have $\varphi_{w_2}(\bar{P}_A) = m_2$. However, the facts $\varphi_{w_2}(\bar{P}_A) = m_2$ and $\varphi_{w_2}(\bar{P}'_{w_2}, \bar{P}_{-w_2}) = m_3$, along with $\bar{P}_{w_2}$'s structure, together imply $\varphi_{w_2}(\bar{P}_A) \mathrel{\bar{P}_{w_2}} \varphi_{w_2}(\bar{P}'_{w_2}, \bar{P}_{-w_2})$, which contradicts the fact that $w_2$ manipulates $\varphi$ at $\bar{P}_A$ via $\bar{P}'_{w_2}$.

				\item Suppose $a = w_3$.
				
				It follows from the construction of $\varphi$ that
				\begin{subequations}\label{equation w3 manipulative}
					\begin{equation}\label{equation w3 manipulative 1}
						\varphi_{w_3}(P_A) = m_2 \mbox{ for all } P_A \in \mathcal{\hat{P}}_A, \mbox{ and}
					\end{equation}
					\begin{equation}\label{equation w3 manipulative 2}
						\varphi_{w_3}(P_A) = D^M_{w_3}(P_A) \mbox{ for all } P_A \in \mathcal{P}_A \setminus \mathcal{\hat{P}}_A.
					\end{equation}
				\end{subequations}
				Using a similar argument as for Case \ref{case m2 manipulator} (and using \eqref{equation w3 manipulative} instead of \eqref{equation m2 manipulative}), it follows that one of $\bar{P}_A$ and $(\bar{P}'_{w_3}, \bar{P}_{-w_3})$ is in $\mathcal{\hat{P}}_A$, and the other is in $\mathcal{P}_A \setminus \mathcal{\hat{P}}_A$. 
				
				Moreover, using a similar argument as for Case \ref{case w2 manipulator} (and using \eqref{equation w3 manipulative} instead of \eqref{equation w2 manipulative}), it follows that $\varphi_{w_3}(\bar{P}_A) \mathrel{\bar{P}_{w_3}} \varphi_{w_3}(\bar{P}'_{w_3}, \bar{P}_{-w_3})$, which contradicts the fact that $w_3$ manipulates $\varphi$ at $\bar{P}_A$ via $\bar{P}'_{w_3}$.

				\item Suppose $a = w_4$.
				
				It follows from the construction of $\varphi$ that
				\begin{subequations}\label{equation w4 manipulative}
					\begin{equation}\label{equation w4 manipulative 1}
						\varphi_{w_4}(P_A) = m_5 \mbox{ for all } P_A \in \mathcal{\tilde{P}}_A, \mbox{ and}
					\end{equation}
					\begin{equation}\label{equation w4 manipulative 2}
						\varphi_{w_4}(P_A) = D^M_{w_4}(P_A) \mbox{ for all } P_A \in \mathcal{P}_A \setminus \mathcal{\tilde{P}}_A.
					\end{equation}
				\end{subequations}
				Using a similar argument as for Case \ref{case m2 manipulator} (and using \eqref{equation w4 manipulative} instead of \eqref{equation m2 manipulative}), it follows that one of $\bar{P}_A$ and $(\bar{P}'_{w_4}, \bar{P}_{-w_4})$ is in $\mathcal{\tilde{P}}_A$, and the other is in $\mathcal{P}_A \setminus \mathcal{\tilde{P}}_A$.
				
				Moreover, using a similar argument as for Case \ref{case w2 manipulator} (and using \eqref{equation w4 manipulative} instead of \eqref{equation w2 manipulative}), it follows that $\varphi_{w_4}(\bar{P}_A) \mathrel{\bar{P}_{w_4}} \varphi_{w_4}(\bar{P}'_{w_4}, \bar{P}_{-w_4})$, which contradicts the fact that $w_4$ manipulates $\varphi$ at $\bar{P}_A$ via $\bar{P}'_{w_4}$.

				\item Suppose $a = w_5$.
				
				It follows from the construction of $\varphi$ that
				\begin{subequations}\label{equation w5 manipulative}
					\begin{equation}\label{equation w5 manipulative 1}
						\varphi_{w_5}(P_A) = m_4 \mbox{ for all } P_A \in \mathcal{\tilde{P}}_A, \mbox{ and}
					\end{equation}
					\begin{equation}\label{equation w5 manipulative 2}
						\varphi_{w_5}(P_A) = D^M_{w_5}(P_A) \mbox{ for all } P_A \in \mathcal{P}_A \setminus \mathcal{\tilde{P}}_A.
					\end{equation}
				\end{subequations}
				Using a similar argument as for Case \ref{case m2 manipulator} (and using \eqref{equation w5 manipulative} instead of \eqref{equation m2 manipulative}), it follows that one of $\bar{P}_A$ and $(\bar{P}'_{w_5}, \bar{P}_{-w_5})$ is in $\mathcal{\tilde{P}}_A$, and the other is in $\mathcal{P}_A \setminus \mathcal{\tilde{P}}_A$.
				
				Moreover, using a similar argument as for Case \ref{case w2 manipulator} (and using \eqref{equation w5 manipulative} instead of \eqref{equation w2 manipulative}), it follows that $\varphi_{w_5}(\bar{P}_A) \mathrel{\bar{P}_{w_5}} \varphi_{w_5}(\bar{P}'_{w_5}, \bar{P}_{-w_5})$, which contradicts the fact that $w_5$ manipulates $\varphi$ at $\bar{P}_A$ via $\bar{P}'_{w_5}$.
				\hfill
				$\Diamond$
			\end{enumerate}
		\end{example}

		Example \ref{example SP but not WGSP} highlights an important observation: not every stable and strategy-proof rule is simply a combination of the matching rules $D^M$ and $D^W$. For instance, in Example \ref{example SP but not WGSP}, consider any preference profile $P_A \in \mathcal{\tilde{P}}_A \setminus \{P^*_A\}$. By construction, $\varphi(P_A) = \mu_2$. However, the matching $\mu_2$ is neither the outcome of the MPDA algorithm nor the WPDA algorithm at $P_A$.

		\section{Structures of single-peaked subdomains under top dominance}\label{appendix domain structure}
		
		\setcounter{equation}{0}
		\setcounter{table}{0}
		\setcounter{theorem}{0}
		\setcounter{definition}{0}
		\setcounter{obs}{0}
		\setcounter{remark}{0}
		\setcounter{example}{0}

		As established by Theorem \ref{theo sp-wgsp classical single peaked}, satisfying top dominance (for at least one side of the market) is both a necessary and sufficient condition for a rich and anonymous (classical) single-peaked subdomain to have stable and strategy-proof matching rules. This naturally leads to the following question: What types of domain structures emerge within the single-peaked subdomains under top dominance? In this section, we partially answer this question by identifying a class of such domains.
		
		Throughout this section, we assume that $\mathcal{P}_A$ is a rich and anonymous single-peaked subdomain and, without loss of generality, that it satisfies top dominance for women. Our goal is to explore the potential structures of $\mathcal{P}_{women}$.
		
		As previously mentioned, one possible structure is $\mathcal{P}_{women}$ consisting of all left-biased single-peaked preferences. Similarly, $\mathcal{P}_{women}$ consisting of all \textit{right-biased single-peaked preferences} is another example. (A single-peaked preference $P \in \mathbb{S}(\prec_M)$ is \textit{\textbf{right-biased}} if for every $m, \tilde{m} \in M \setminus \{\tau(P)\}$, $m \prec_M \tau(P) \prec_M \tilde{m}$ implies $\tilde{m} \mathrel{P} m$.) Building on these two examples, we identify a class of structures for $\mathcal{P}_{women}$ that ensures $\mathcal{P}_A$ to satisfy top dominance for women. We use the following notions and notations to do so.
		
		For two distinct men $m, m' \in M$ with $m \prec_M m'$, we say that $m$ is \textit{to the left} of $m'$, and $m'$ is \textit{to the right} of $m$. For every man $m \in M$, let $M^L_m = \{m' \in M \mid m' \preceq m\}$ be the set consisting of $m$ and all men to the left of him. Similarly, let $M^R_m = M \setminus M^L_m$ be the set consisting of all men to the right of $m$. 
		Given a man $m \in M$, a single-peaked preference $P \in \mathbb{S}(\prec_M)$ is said to be \textbf{\textit{$(M^L_m\mbox{-left}, M^R_m\mbox{-right})$-biased}} if
		\begin{enumerate}[(i)]
			\item it is left-biased when $\tau(P) \in M^L_m$, and
			
			\item it is right-biased when $\tau(P) \in M^R_m$.
		\end{enumerate}
		
		In essence, the concept of $(M^L_m\mbox{-left}, M^R_m\mbox{-right})$-biasedness generalizes both left-biasedness and right-biasedness. Specifically, when man $m$ is the right-most man (according to $\prec_M$), $(M^L_m\mbox{-left}, M^R_m\mbox{-right})$-biasedness reduces to left-biasedness; when $m$ is the left-most man, $(M^L_m\mbox{-left}, M^R_m\mbox{-right})$-biasedness boils down to right-biasedness.
		
		We now present the main result of this section, which identifies a class of structures for $\mathcal{P}_{women}$ that ensures $\mathcal{P}_A$ to satisfy top dominance for women.

		\begin{proposition}\label{proposition single-peaked top dominance construction}
			Suppose $\mathcal{P}_A$ is a rich and anonymous (classical) single-peaked subdomain. Fix $m^* \in M$. If $\mathcal{P}_{women}$ consists of all $(M^L_{m^*}\mbox{-left}, M^R_{m^*}\mbox{-right})$-biased single-peaked preferences, then $\mathcal{P}_A$ satisfies top dominance for women.
		\end{proposition}

		\begin{proof}[\textbf{Proof of Proposition \ref{proposition single-peaked top dominance construction}}]
			Suppose $\mathcal{P}_{women}$ consists of all $(M^L_{m^*}\mbox{-left}, M^R_{m^*}\mbox{-right})$-biased single-peaked preferences. 
			Assume for contradiction that $\mathcal{P}_A$ does not satisfy top dominance for women. Since $\mathcal{P}_A$ is anonymous and does not satisfy top dominance for women, there exist distinct $m, \tilde{m}, m' \in M$ and two preferences $P_1, P_2 \in \mathcal{P}_{women}$ such that 
			\begin{subequations}\label{equation NTD}
				\begin{equation}\label{equation NTD 1}
					m \mathrel{P_1} \tilde{m} \mathrel{P_1} m', \mbox{ and}
				\end{equation}
				\begin{equation}\label{equation NTD 2}
					m \mathrel{P_2} m' \mathrel{P_2} \tilde{m}.
				\end{equation}
			\end{subequations}
			Since $\mathcal{P}_A$ is a single-peaked subdomain, \eqref{equation NTD} implies 
			\begin{equation*}
				\tilde{m} \prec_M m \prec_M m' \hspace{2 mm} \mbox{ or } \hspace{2 mm} m' \prec_M m \prec_M \tilde{m}.
			\end{equation*}
			We distinguish the following two cases.\medskip

			\noindent\textbf{\textsc{Case} 1}: Suppose $\tilde{m} \prec_M m \prec_M m'$.
			
			Since $\tilde{m} \prec_M m \prec_M m'$, \eqref{equation NTD 1}, along with single-peakedness of $P_1$, implies $\tilde{m} \prec_M \tau(P_1) \prec_M m'$. This, together with \eqref{equation NTD 1} and the fact that $P_1$ is $(M^L_{m^*}\mbox{-left}, M^R_{m^*}\mbox{-right})$-biased, implies
			\begin{equation}\label{equation NTD case 1 1}
				\tau(P_1) \preceq_M m^*.
			\end{equation}
			Furthermore, since $P_1$ is a $(M^L_{m^*}\mbox{-left}, M^R_{m^*}\mbox{-right})$-biased single-peaked preference, \eqref{equation NTD case 1 1} essentially means $P_1$ is a left-biased single-peaked preference. Because of this, and since $\tilde{m} \prec_M m \prec_M m'$, by \eqref{equation NTD 1}, we have
			\begin{equation}\label{equation NTD case 1 2}
				m \preceq_M \tau(P_1).
			\end{equation}
			
			Similarly, since $\tilde{m} \prec_M m \prec_M m'$, \eqref{equation NTD 2}, along with single-peakedness of $P_2$, implies $\tilde{m} \prec_M \tau(P_2) \prec_M m'$. This, together with \eqref{equation NTD 2} and the fact that $P_2$ is $(M^L_{m^*}\mbox{-left}, M^R_{m^*}\mbox{-right})$-biased, implies
			\begin{equation}\label{equation NTD case 1 3}
				m^* \prec_M \tau(P_2).
			\end{equation}
			Furthermore, since $P_2$ is a $(M^L_{m^*}\mbox{-left}, M^R_{m^*}\mbox{-right})$-biased single-peaked preference, \eqref{equation NTD case 1 3} essentially means $P_2$ is a right-biased single-peaked preference. Because of this, and since $\tilde{m} \prec_M m \prec_M m'$, by \eqref{equation NTD 2}, we have
			\begin{equation}\label{equation NTD case 1 4}
				\tau(P_2) \preceq_M m.
			\end{equation}
			
			However, \eqref{equation NTD case 1 4}, \eqref{equation NTD case 1 2}, and \eqref{equation NTD case 1 1} together imply $\tau(P_2) \preceq_M m^*$, a contradiction to \eqref{equation NTD case 1 3}.\medskip

			\noindent\textbf{\textsc{Case} 2}: Suppose $m' \prec_M m \prec_M \tilde{m}$.
			
			Since $m' \prec_M m \prec_M \tilde{m}$, \eqref{equation NTD 1}, along with single-peakedness of $P_1$, implies $m' \prec_M \tau(P_1) \prec_M \tilde{m}$. This, together with \eqref{equation NTD 1} and the fact that $P_1$ is $(M^L_{m^*}\mbox{-left}, M^R_{m^*}\mbox{-right})$-biased, implies
			\begin{equation}\label{equation NTD case 2 1}
				m^* \prec_M \tau(P_1).
			\end{equation}
			Furthermore, since $P_1$ is a $(M^L_{m^*}\mbox{-left}, M^R_{m^*}\mbox{-right})$-biased single-peaked preference, \eqref{equation NTD case 2 1} essentially means $P_1$ is a right-biased single-peaked preference. Because of this, and since $m' \prec_M m \prec_M \tilde{m}$, by \eqref{equation NTD 1}, we have
			\begin{equation}\label{equation NTD case 2 2}
				\tau(P_1) \preceq_M m.
			\end{equation}
			
			Similarly, since $m' \prec_M m \prec_M \tilde{m}$, \eqref{equation NTD 2}, along with single-peakedness of $P_2$, implies $m' \prec_M \tau(P_2) \prec_M \tilde{m}$. This, together with \eqref{equation NTD 2} and the fact that $P_2$ is $(M^L_{m^*}\mbox{-left}, M^R_{m^*}\mbox{-right})$-biased, implies
			\begin{equation}\label{equation NTD case 2 3}
				\tau(P_2) \preceq_M m^*.
			\end{equation}
			Furthermore, since $P_2$ is a $(M^L_{m^*}\mbox{-left}, M^R_{m^*}\mbox{-right})$-biased single-peaked preference, \eqref{equation NTD case 2 3} essentially means $P_2$ is a left-biased single-peaked preference. Because of this, and since $m' \prec_M m \prec_M \tilde{m}$, by \eqref{equation NTD 2}, we have
			\begin{equation}\label{equation NTD case 2 4}
				m \preceq_M \tau(P_2).
			\end{equation}
			
			However, \eqref{equation NTD case 2 2}, \eqref{equation NTD case 2 4}, and \eqref{equation NTD case 2 3} together imply $\tau(P_1) \preceq_M m^*$, a contradiction to \eqref{equation NTD case 2 1}.\medskip
			
			Since Cases 1 and 2 are exhaustive, this completes the proof of Proposition \ref{proposition single-peaked top dominance construction}.
		\end{proof}

		The converse of Proposition \ref{proposition single-peaked top dominance construction} does not hold in general. Beyond the structures identified in Proposition \ref{proposition single-peaked top dominance construction}, other configurations for $\mathcal{P}_{women}$ can also ensure that $\mathcal{P}_A$ satisfies top dominance for women. The following example illustrates such a case.

		\begin{example}
			Suppose $M = \{m_1, m_2, m_3, m_4\}$ with a prior (linear) ordering $\prec_M$ such that $m_1 \prec_M m_2 \prec_M m_3 \prec_M m_4$. Consider a rich and anonymous single-peaked subdomain $\mathcal{P}_A$ such that 
			\begin{equation*}
				\mathcal{P}_{women} = \{m_1 m_2 m_3 m_4, m_2 m_3 m_1 m_4, m_3 m_2 m_1 m_4, m_4 m_3 m_2 m_1\}.
			\end{equation*}
			
			It is straightforward to verify that $\mathcal{P}_A$ satisfies top dominance for women. However, $P_{women}$ has a structure that is not identified by Proposition \ref{proposition single-peaked top dominance construction}. This is because the preference $m_2 m_3 m_1 m_4$ is neither left-biased nor right-biased and, consequently, does not exhibit $(M^L_m\mbox{-left}, M^R_m\mbox{-right})$-biasedness for any $m \in M$.
			\hfill
			$\Diamond$
		\end{example}

	\end{appendices}

	\section*{Declarations}

	\paragraph{Declaration of competing interest}
	
	The author has no competing interests to declare that are relevant to the content of this article.

	\paragraph{Data availability}
	
	No data was used for the research described in the article.

	\bibliographystyle{plainnat}
	\setcitestyle{numbers}
	\bibliography{mybib}

\end{document}